\documentclass[aps,prl,twocolumn,superscriptaddress,longbibliography,amsmath,amssymb]{revtex4-2}
\usepackage{xcolor}
\usepackage{graphicx}
\usepackage{units}
\usepackage[frenchb]{babel}
\usepackage[utf8]{inputenc}
\usepackage{amssymb,amsmath}
\usepackage{tabularx}
\usepackage{multirow}
\usepackage{booktabs}
\usepackage{colortbl}
\usepackage{tabularx}
\usepackage{color}
\usepackage{systeme}

\begin{document}
\title{Exciton self-trapping in twisted hexagonal boron nitride homostructures}

\author{S\'ebastien Roux}
\affiliation{Universit\'e Paris-Saclay, ONERA-CNRS, LEM, 92320 Ch\^atillon, France}
\affiliation{Universit\'e Paris-Saclay, UVSQ, CNRS, GEMaC, 78000, Versailles, France}
\email{sroux@insa-toulouse.fr}

\author{Christophe Arnold}
\affiliation{Universit\'e Paris-Saclay, UVSQ, CNRS, GEMaC, 78000, Versailles, France}
\author{Etienne Carr\'e}
\affiliation{Universit\'e Paris-Saclay, ONERA-CNRS, LEM, 92320 Ch\^atillon, France}
\affiliation{Universit\'e Paris-Saclay, UVSQ, CNRS, GEMaC, 78000, Versailles, France}
\author{Alexandre Plaud}
\affiliation{Universit\'e Paris-Saclay, ONERA-CNRS, LEM, 92320 Ch\^atillon, France}
\affiliation{Universit\'e Paris-Saclay, UVSQ, CNRS, GEMaC, 78000, Versailles, France}
\author{Lei Ren}
\affiliation{Universit\'e de Toulouse, INSA-CNRS-UPS, LPCNO, 135 Av. Rangueil, 31077 Toulouse, France}
\author{Fr\'ed\'eric Fossard}
\affiliation{Universit\'e Paris-Saclay, ONERA-CNRS, LEM, 92320 Ch\^atillon, France}
\author{Nicolas Horezan}
\affiliation{Universit\'e Paris-Saclay, ONERA, DMAS, 92320 Ch\^atillon, France}
\author{Eli Janzen}
\affiliation{ Tim Taylor Department of Chemical Engineering, Kansas State University Manhattan, KS 66506, USA}
\author{James H. Edgar}
\affiliation{ Tim Taylor Department of Chemical Engineering, Kansas State University Manhattan, KS 66506, USA}
\author{Camille Maestre}
\affiliation{Laboratoire des Multimat\'eriaux et Interfaces, UMR CNRS 5615, Univ Lyon \\ Universit\'e Claude Bernard Lyon 1, F-69622 Villeurbanne, France}
\author{B\'erang\`ere Toury}
\affiliation{Laboratoire des Multimat\'eriaux et Interfaces, UMR CNRS 5615, Univ Lyon \\ Universit\'e Claude Bernard Lyon 1, F-69622 Villeurbanne, France}
\author{Catherine Journet}
\affiliation{Laboratoire des Multimat\'eriaux et Interfaces, UMR CNRS 5615, Univ Lyon \\ Universit\'e Claude Bernard Lyon 1, F-69622 Villeurbanne, France}
\author{Vincent Garnier}
\affiliation{Laboratoire MATEIS, UMR CNRS 5510, Univ Lyon, INSA Lyon,  F-69621 Villeurbanne, France}
\author{Philippe Steyer}
\affiliation{Laboratoire MATEIS, UMR CNRS 5510, Univ Lyon, INSA Lyon,  F-69621 Villeurbanne, France}
\author{Takashi Taniguchi}
\affiliation{Research Center for Materials Nanoarchitectonics, National Institute for Materials Science,  1-1 Namiki, Tsukuba 305-0044, Japan}
\author{Kenji Watanabe}
\affiliation{Research Center for Electronic and Optical Materials, National Institute for Materials Science, 1-1 Namiki, Tsukuba 305-0044, Japan}
\author{Cédric Robert}
\affiliation{Universit\'e de Toulouse, INSA-CNRS-UPS, LPCNO, 135 Av. Rangueil, 31077 Toulouse, France}
\author{Xavier Marie}
\affiliation{Universit\'e de Toulouse, INSA-CNRS-UPS, LPCNO, 135 Av. Rangueil, 31077 Toulouse, France}
\author{Fran\c cois Ducastelle}
\affiliation{Universit\'e Paris-Saclay, ONERA-CNRS, LEM, 92320 Ch\^atillon, France}
\author{Annick Loiseau}
\affiliation{Universit\'e Paris-Saclay, ONERA-CNRS, LEM, 92320 Ch\^atillon, France}
\author{Julien Barjon}
\affiliation{Universit\'e Paris-Saclay, UVSQ, CNRS, GEMaC, 78000, Versailles, France}

\date{\today}
\begin{abstract}
One of the main interests of 2D materials is their ability to be assembled with many degrees of freedom for tuning and manipulating excitonic properties. There is a need to understand how the structure of the interfaces between atomic layers influences exciton properties. Here we use cathodoluminescence and time-resolved cathodoluminescence experiments to study how excitons interact with the interface between two twisted hexagonal boron nitride (hBN) crystals with various angles. An efficient capture of free excitons by the interface is demonstrated, which leads to a population of long-lived and interface-localized (2D) excitons. Temperature dependent experiments indicate that for high twist angles, these excitons localized at the interface further undergo a self-trapping. It consists in a distortion of the lattice around the exciton on which the exciton traps itself. Our results suggest that this exciton-interface interaction causes the broad 4-eV optical emission of highly twisted hBN-hBN structures. Exciton self-trapping is finally discussed as a common feature of sp$^{2}$ hybridized boron nitride polytypes and nanostructures due to the ionic nature of the B-N bond and the small size of their excitons.
\end{abstract}

\maketitle 
 
\subsection{I - Introduction}
The conception of 2D material heterostructures (h2D) benefits from a large number of degrees of freedom in the choice of atomic layers and the way they are stacked, creating structural singularities at the interfaces between the layers \cite{Geim2013, Sierra2021, Lau2022, Gibertini2019}. Taking advantage of these capabilities allows to control and manipulate the properties of excitons: the efficiency of their radiative recombination \cite{Fang2019, Choi2021, Rivera2015, Ovesen2019, Miller2017, Yuan2020}, the properties of their emission \cite{Cadiz2017}, their diffusion length, their valley and/or spin coherence \cite{Rivera2016, Kim2017} and the dielectric screening between the electron and the hole \cite{Carre2022,Latini2015, Lin2014, Hsu2019}. Finally, excitons can interact with Moiré-superpotentials, which tune their properties and motion \cite{Alexeev2019, Jin2019, Seyler2019, Tran2019, Baek2020, Brem2020, Gisbert2020}. Thus, h2Ds offer an ideal platform for creating novel electronic devices using exciton fluxes (excitronics) \cite{Unuchek2018}, or using their spin/valley indices (valleytronics and spintronics) \cite{Unuchek2019}. 
Hexagonal boron nitride (hBN) is present in most h2D since it is the best insulating material for use as a substrate or as a capping layer for other 2D materials such as transition metal dichalcogenides (TMDs), 2D magnets \cite{Gong2019} and graphene \cite{Wang2010, Dean2010, Cadiz2017, Arora2019, Huang2018, Kim2019, Carre2022}. Because hBN plays such a key role in h2D-based devices, it is crucial to understand the surface and interface effects associated with it. 

The simplest hBN-based h2D is the twisted hBN-hBN homostructure, which consists of two hBN crystals with different in-plane crystal orientations. These structures exhibit many properties that are not present in single hBN crystals. At small twist angles, triangular ferroelectric domains appear at the interface \cite{Yasuda2021, Woods2021}, as well as drastic changes in the electronic properties, such as in some cases, a spatial separation of the electron and hole wave functions within the Moiré supercell \cite{Zhao2020, Zhao2021}. At high twist angles, hBN-hBN homostructure exhibits intense second harmonic generation, which could be modulated by the twist angle \cite{Yao2021}. Finally, a new 4-eV luminescence signal has recently been discovered in cathodoluminescence (CL) on twisted hBN-hBN structures. 

\begin{figure}[h!]
\includegraphics[scale=0.55]{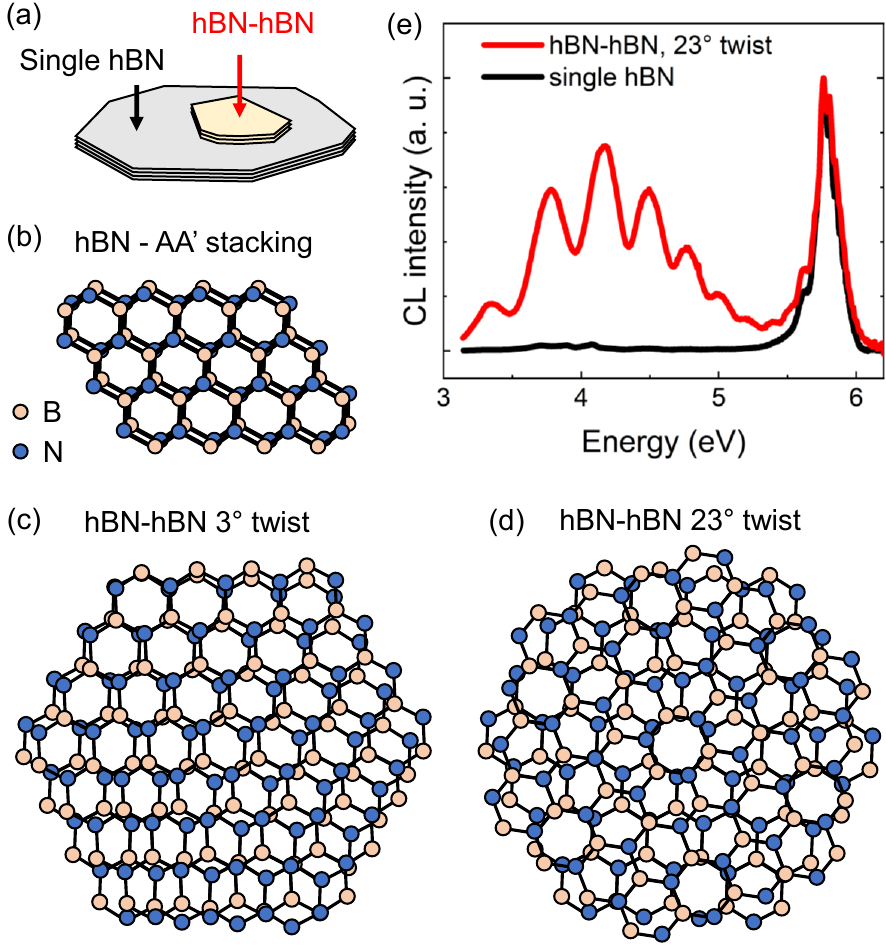}
\caption{(a) Sketch of a hBN-hBN structure. Top views of (b) the AA' stacking arrangement of hBN, (c) a 3°-twisted hBN-hBN interface and (d) a 23°-twisted hBN-hBN interface. (e) CL spectra measured on a single bulk hBN crystal (the bottom part alone), in black, and on the 23° hBN-hBN stack, in red. The luminescence is modulated by Fabry-Perot interferences related to the thinness of the crystals.
}
\label{F0}
\end{figure}

Twisted hBN-hBN structures are composed of two stacked multilayer hBN flakes, as shown in Fig.~\ref{F0} (a). On the one hand, the single flakes are arranged in the AA' stacking configuration (Fig.~\ref{F0} (b)), with high symmetry and exclusively out-of-plane hetero $\pi^{B-N}$ bonds. On the other hand, the twisted interface presents a variety of out-of-plane bonds, including homo $\pi^{B-B}$ and $\pi^{N-N}$ bonds (Fig.~\ref{F0} (c-d)). For small twist angles, a macroscopic moiré superstructure appears, with $\pi^{B-B}$ and $\pi^{N-N}$ localized at specific positions of the interface. For large twist angles, close to the 30° twist quasi-crystal limit, atomic orbitals vary locally from site to site, and homo $\pi^{B-B}$ and $\pi^{N-N}$ bonds are densely distributed all over the interface. The luminescence of single flakes is dominated by the free exciton emission, while highly twisted hBN-hBN structures exhibit a new emission, shown in Fig.~\ref{F0} (e), characterized by a large linewidth (2~eV) and a maximum intensity at 4~eV, i.e. 2~eV below the hBN gap \cite{Plaud2020, Lee2021}. The origin of this optical emission, its 2-eV redshift, and its relation to the structure of the interface are debated in the literature. First, a giant exciton trapping at the interface moiré was proposed \cite{Lee2021, Li2022}. However, this is not consistent with theoretical studies, which estimate the depth of the interface energy well to be only a few hundreds of meV for excitons \cite{Su2022, Latil2023}. Therefore, another group has discarded its excitonic nature and rather attributed this emission to deep defects near the interface \cite{Su2022}.

In this paper we propose an alternative explanation with the occurrence of exciton self-trapping at the interface of the homostructure. The phenomenon of self-trapping of an electron, hole, or exciton was predicted theoretically in the 1930s \cite{Landau1933, Frenkel1936} and demonstrated experimentally in alkali halide crystals as early as the 1960s \cite{Castner1957, Kabler1964, Murray1965, Ramamurti1966, Kink1967}. Self-trapping results from a local deformation of the crystal lattice around a charged particle (or dipole) where the particle (or dipole) is trapped. The particle (or dipole) induces the lattice deformation, hence the term "self-trapped state". This state is a kind of polaron: a charge associated with the deformation cloud it induces around itself. If the coupling with the deformation mode is strong enough, the self-trapped state is more energetically stable than the Bloch state of the free particle in the undeformed lattice \cite{Landau1933}. hBN is a good candidate for exciton self-trapping since it is a highly ionic material with a very compact exciton \cite{Roux2021}. It makes excitons and atomic B-N bonds dipoles of similar size, which is expected to favor exciton-lattice interactions.

Today, the puzzle for understanding the nature and the origin of this new optical emission is incomplete. In this work, we present a deterministic approach to elucidate the interplay between twist angles, defects and excitons using CL and time-resolved CL (TRCL). To this end, sixteen hBN-hBN structures with different twist angles and fabricated from different hBN crystal sources are studied in CL and TRCL, at room temperature and as a function of temperature. We show how the whole set of experimental data supports an exciton self-trapping mechanism occurring at the interface of hBN-hBN homostructures.

\subsection{II – Experiments}
All CL experiments were performed using a JEOL7001F scanning electron microscope (SEM) equipped with a Horiba Jobin-Yvon CL system optimized for UV detection, as described in detail in reference \cite{Schue2019}. Secondary electrons are detected by a Everhart-Thornley detector (SE detector). The SEM image corresponds to the intensity of the signal collected by the SE detector as a function of the position of the focused electron beam, with an ultimate spatial resolution of 1.2~nm.
CL images are measured by a photomultiplier (Hamamatsu HJY model - cooled to -30°C by Peltier effect) as the electron beam scans the selected area of the sample. The signal from the SE detector and the intensity collected by the photomultiplier are recorded simultaneously. In this way, SEM and CL images of the same area are generated, which allows to correlate the emission of a CL signal with the topography of the sample. The CL image can be panchromatic (total intensity, unfiltered) or monochromatic, filtered in energy by a monochromator equiped with diffraction gratings. CL spectra are recorded with a CCD camera. 

Thanks to a careful calibration of the detection system, the CL signal provides the absolute intensity of light emission from the sample \cite{Schue2019}. As a result, the internal quantum efficiency ($IQE$) of a luminescence signal can be evaluated from a CL experiment. The $IQE$ corresponds to the ratio between the rate of photons emitted inside the sample and the rate of electron-hole pairs generated by the electron bombardment. Details of the $IQE$ measurements for the considered signal are given in the Appendix A.
Finally, TRCL experiments are performed using a custom-built beam-blanking device mounted on the SEM column. The overall time resolution of the TRCL set-up is measured equal to 100~ps (details of the setup are presented in ref. \cite{Roux2021}).

To study the emission of twisted hBN-hBN structures, we decided to vary a large number of parameters in the fabrication of the homostructures: the twist angle, the thickness of the crystal flakes in stack, but also the quality of the starting hBN crystals and the way the two crystal flakes are assembled. hBN crystals with natural boron isotope content were grown by three different processes: an Atmospheric Pressure High Temperature (APHT) process \cite{Kubota2007, Hoffman2014, Liu2018, Li2020} from Ni/Cr solvent, a High-Pressure High Temperature (HPHT) route \cite{Watanabe2004, Taniguchi2007}, and a Polymer Derived Ceramic (PDC) method \cite{Li2020b, Maestre2022}. hBN from the HPHT method is recognized in the 2D materials scientific community as the reference hBN crystals. In a previous study, a quantitative benchmarking of the respective quality, in terms of defect density, of these different hBN sources was performed based on the measured free exciton lifetimes and correlated to the electron mobility in graphene/hBN h2Ds \cite{Roux2021, Ouaj2024}. The values of the lifetime and $IQE$ of the free exciton measured on bulk crystals from the different hBN sources are given in Tab.~\ref{T1}.

\begin{table}[h]	
\begin{ruledtabular}
			\begin{tabular}{l|ccrr}
\multirow{2}{*}{Sample}  & Free exciton & $\tau$ (ns) & Top hBN  & Bottom hBN  \\
 & $IQE$ (\%) \cite{Roux2021} & \cite{Roux2021} & (nm) & (nm) \\

				\hline
HPHT-11° & \multirow{5}{*}{18} & \multirow{5}{*}{4.2} & 260 & 730  \\
HPHT-14° & & & 15 & 290 \\
HPHT-15° & & & 17 & 990 \\
HPHT-23° & & & 9 & 220\\
HPHT-27° & & & 13 & 990\\

				\hline
APHT-3° & \multirow{5}{*}{4.2} & \multirow{5}{*}{1.0} & 60 & 40  \\
APHT-4° & & & 67 & 37 \\
APHT-10° & & & 84 & 38 \\
APHT-13° & & & 100 & 60\\
APHT-29° & & & 20 & 50\\

				\hline
PDC-2° & \multirow{6}{*}{1.7} & \multirow{6}{*}{0.43} & 27 & 46  \\
PDC-13° & & & 68 & 46 \\
PDC-16° & & & 310 & 125 \\
PDC-18° & & & 310 & 370\\
PDC-19° & & & 44 & 46\\
PDC-26° & & & 102 & 46\\
			\end{tabular}
			\end{ruledtabular}
			\caption{Presentation of the 16 hBN-hBN samples of the present study and the respective thickness of top and bottom flakes. The $IQE$ and the lifetime ($\tau$) of the free exciton measured on bulk crystals from the same sources \cite{Roux2021} are indicated to quantitatively compare the quality of the samples.}
			\label{T1}			
		\end{table}%

To fabricate twisted homostructures made of HPHT crystals, the bottom hBN layer is  exfoliated with a PDMS stamp and deposited onto a Si/SiO$_{2}$ substrate \cite{Castellanos2014}. The second hBN flake is similarly exfoliated, and deposited over the bottom hBN under a microscope. The samples are then annealed at 150°C for 1.5 hours. With this method, one atomic layer of the twisted interface was in contact with the PDMS.

For twisted homostructures of APHT and PDC crystals, the bottom layer is peeled off using Scotch tape on a Si/SiO$_{2}$ substrate \cite{Novoselov2004}, while the top layer is peeled off using PDMS \cite{Castellanos2014} and deposited over the bottom hBN flake under the microscope. This method ensures that the two atomic layers of the hBN-hBN interface have never been in contact with either the PDMS or the Scotch tape and does not involve annealing. All the procedures, for HPHT, APHT and PDC samples were performed in a controlled atmosphere glove box.

The twist angle between the two hBN flakes is measured with diffraction by electron backscatter diffraction (EBSD) or by  electron channeling patterns (ECP) as shown in Fig.~\ref{F1}. Details of the twist angle measurement are presented in the Appendix B. The samples are named according to the crystal synthesis method from which they are issued (HPHT, APHT or PDC) and the twist angle. Tab.~\ref{T1} lists the different samples and the respective thickness of the top and bottom parts measured by Atomic Force Microscopy (AFM).

\begin{figure}[h!]
\includegraphics[scale=0.4]{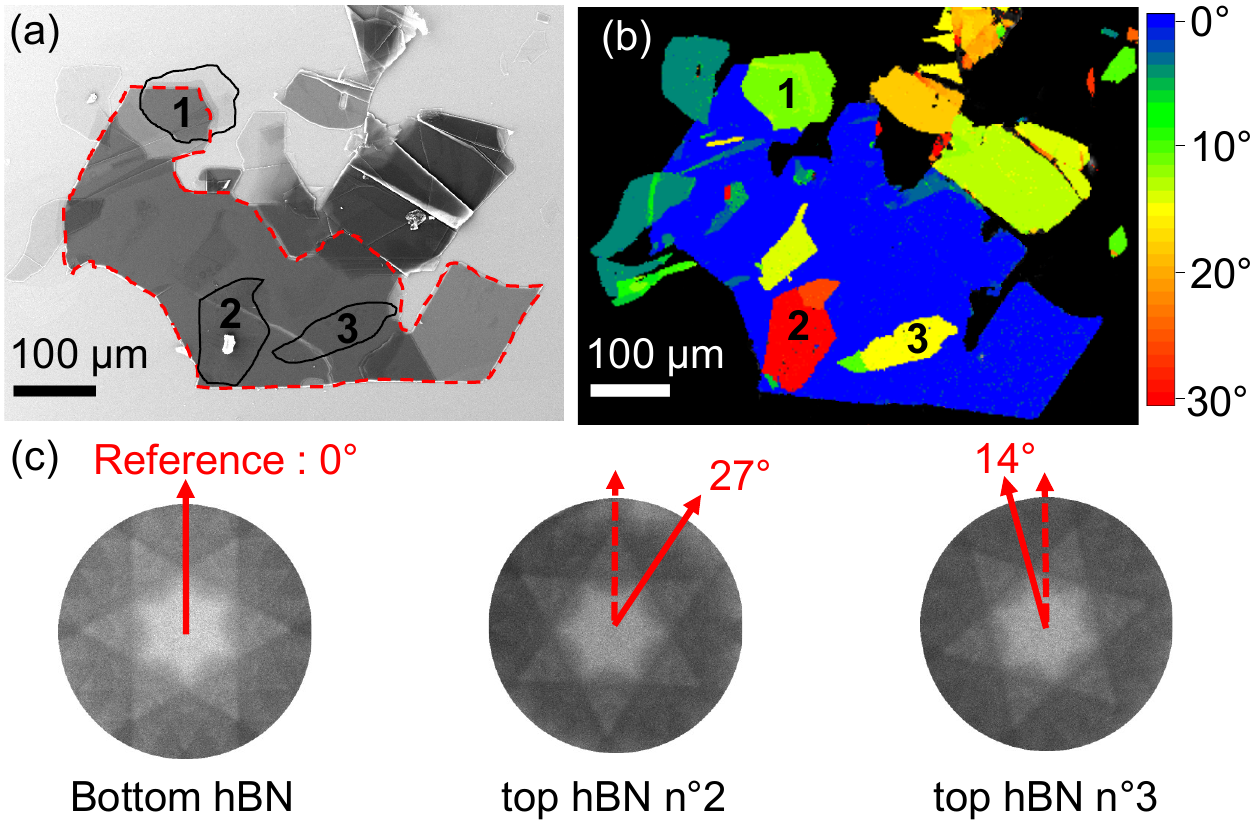}
\caption{(a) SEM image of 3 hBN-hBN homostructures fabricated from an HPHT bulk crystal. The lower hBN flake is indicated by red dotted lines. The upper hBN flakes are numbered from 1 to 3 and surrounded by solid black lines. (b) EBSD mapping of the in-plane crystal orientation of the hBN flakes. The orientation of the bottom hBN crystal is taken as the reference for evaluating the twist angle of the top hBN. (c) ECP recorded on the lower hBN crystal and upper hBN crystals n° 2 and n° 3.
}
\label{F1}
\end{figure}

\subsection{III - Evidence of the excitonic origin of the broad 4-eV optical emission}
\begin{figure}[h!]
\includegraphics[scale=0.32]{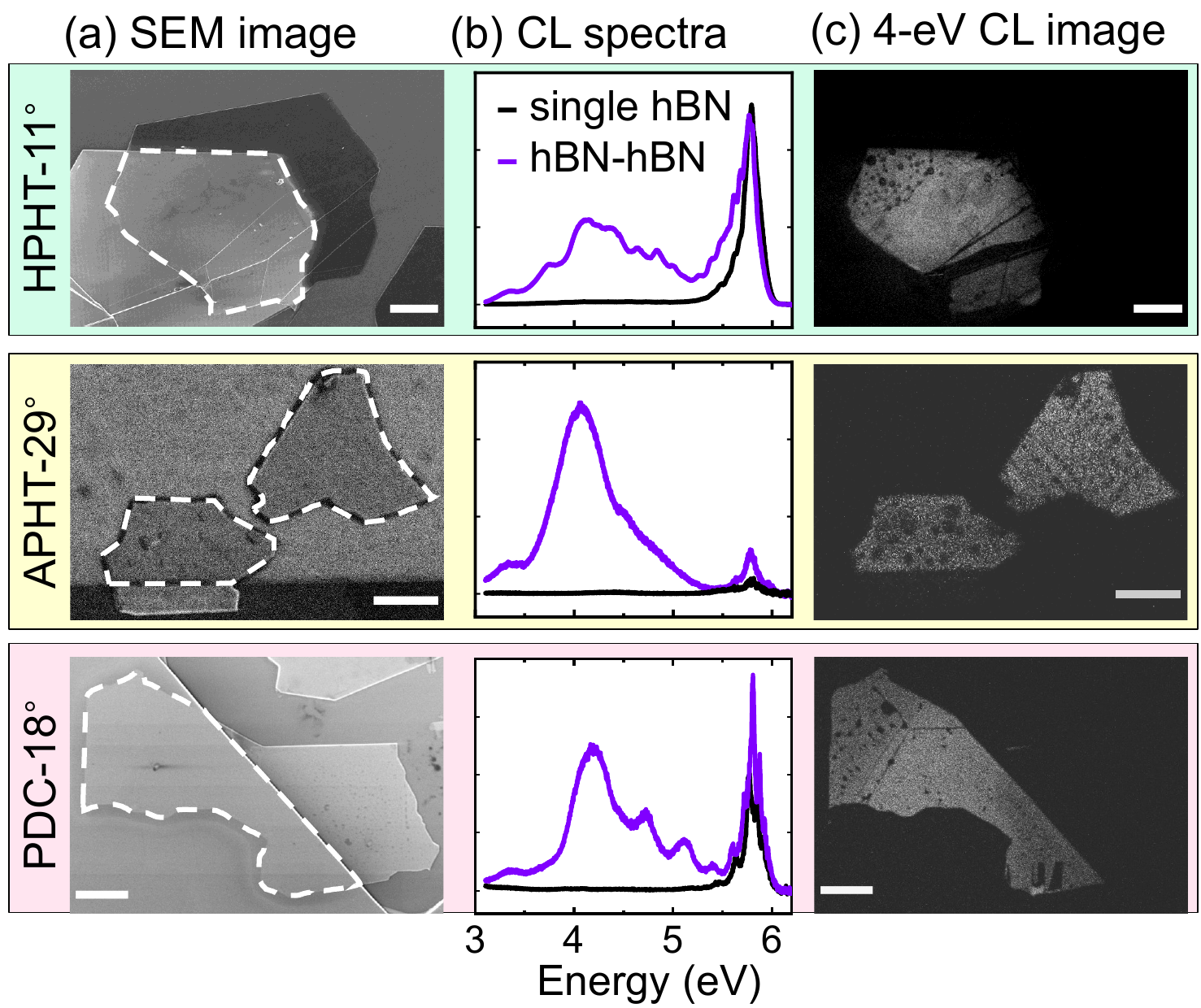}
\caption{(a) SEM images of the twisted homostructures, outlined with a dotted line; (b) CL spectra measured on the bottom crystal alone (in black) and on the homostructure (in purple) at 300~K; (c) CL images filtered at 4.1$\pm$0.1~eV at 300~K, on HPHT-11° (5~kV, 27~pA), APHT-29° (3~kV, 280~pA), and PDC-18° (5~kV, 340~pA) samples. Scale bars are 10~µm.
}
\label{F2}
\end{figure}

All hBN-hBN samples were first analyzed using continuous excitation CL at room temperature. The first observation is that the 4-eV luminescence mainly appears in homostructures with high twist angles (Fig.~\ref{F2}). At low twist, such as the PDC-2°, APHT-3° and APHT-4° samples, no clear luminescence signal could be detected in the 3-5~eV range, confirming a recent study \cite{Lee2021}. We conclude that the 4-eV luminescence becomes truly significant at angles larger than 5°. A spectrum measured on a low angle hBN-hBN structure is presented in the Appendix C. Spectra measured on 3 highly twisted hBN-hBN homostructures (purple) and on a single hBN crystal as a reference (black) are shown in Fig.~\ref{F2}(b). The 4-eV emission is modulated by Fabry-Pérot interferences whose periodicity is consistent with the thickness of the flakes (see Tab.~\ref{T1}). CL images filtered at 4~eV are shown in Fig.~\ref{F2}(c). They clearly demonstrate that the 4~eV cathodoluminescence component is due to the presence of the twisted hBN-hBN interface.

The full set of spectra and CL images recorded for the 16 structures investigated in this work are provided in the Supplementary Information. The signals are comparable for all samples regardless of hBN source and flake thickness. They show common features in line with previous observations \cite{Lee2021, Li2022, Su2022}: intense, 2-eV broadening and 2-eV redshift with respect to the 6.25-eV hBN bandgap \cite{Schue2019}.

In summary of these first observations, the emission phenomenon is very robust and appears unconditionally as soon as the twist angle is large, as already reported in Ref. \cite{Lee2021} for instance. As it does not depend on the differences in nature and density of defects that present the three hBN sources used to fabricate our homostructures, the emission seems to be intrinsic to the presence of the interface itself rather than related to extrinsic defects. This raises the question of its origin, and more specifically, the interplay between the exciton and the twisted-interface structure, which we will explore next using TRCL. 

\begin{figure}[h!]
\includegraphics[scale=0.24]{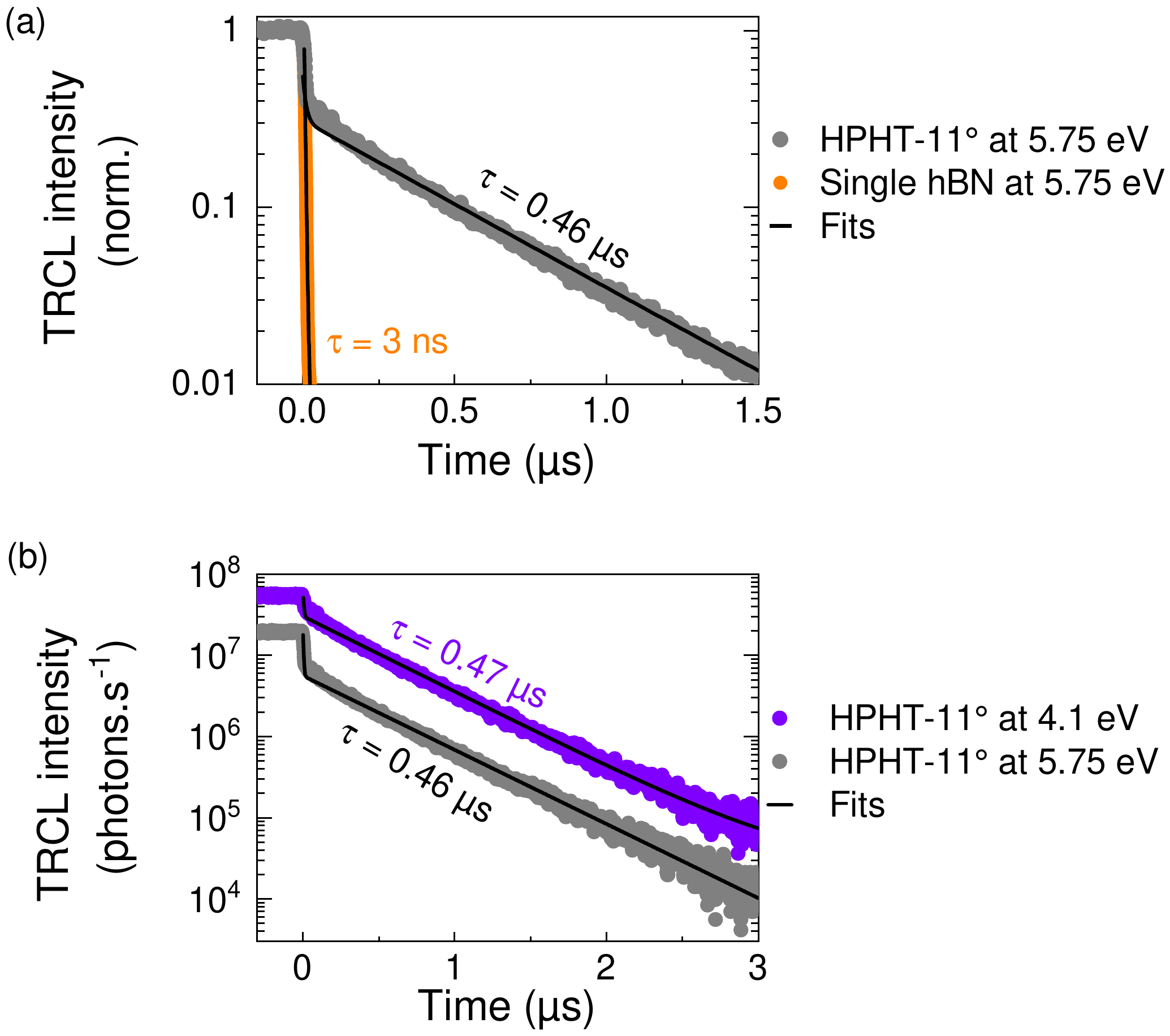}
\caption{(a) Decay of the free exciton luminescence, filtered at 5.75$\pm$0.2~eV, on the bottom hBN alone (gray) and on the hBN-hBN twisted structure HPHT-11° (orange), after interrupting the TRCL excitation at initial time (5~kV, 27~pA, spot size of 10~µm). Bi-exponential decay functions are used for the fits. (b) Decay of the free exciton luminescence (black), and of the the 4-eV emission (purple), filtered at 4.1$\pm$0.1~eV, after interruption of steady excitation at t=0, on HPHT-11° homostructure (5~kV, 27~pA, 10~µm excitation spot diameter). Intensities during steady state excitation (t $<$ 0) are normalized by the integrated intensity of the bands recorded on continuous CL spectra. Experiments done at 300~K.
}
\label{F3}
\end{figure}

The goal of the TRCL study is to identify the role of bulk free excitons in the process leading to the 4-eV luminescence. Figure~\ref{F3}(a) compares the decay of the free exciton luminescence at 5.75~eV in the HPHT-11° homostructure with the decay obtained on the bottom hBN crystal alone. A drastic change is observed. For the crystal alone, the decay is dominated by a short 3~ns component, typical of the TRCL decay of the free exciton in a single bulk HPHT-hBN crystal \cite{Roux2021}. For the homostructure, the fast initial component is similar but a long component of 0.46~µs appears with a significant intensity (30\% on the HPHT-11° sample). This long time constant is much larger than the radiative lifetime of the free exciton measured at 27~ns \cite{Roux2021}, which requires an incoming flux of free excitons that persists after the primary excitation has stopped. We conclude that the long lasting free exciton emission is fed by the detrapping of excitons from the interface, revealing a balance between a population of interface excitons and the population of free excitons in the hBN volume based on a trapping-detrapping process.

As shown in Appendix D, the temporal decay of the 4-eV luminescence is energy-independent in the 3-5 eV energy range. Figure~\ref{F3}(b) further illustrates that its decay time matches the long-lived component of the free exciton emission. Without going into the quantitative analysis (presented later), it demonstrates that the broad 4-eV luminescence is excited by the long-lived and interface-localized excitons. This implies that the 2 eV broadening likely does not arise from energy dispersion of an ensemble of localized quantum emitters. Instead, it appears to result from the radiative recombination process of interface excitons. Still, the radiative process has to be identified, which is the subject of the next parts.

\subsection{IV - Power dependence and internal quantum efficiency of the 4-eV emission.}
During CL experiments, the electron beam excitation is spread in depth, so that the hBN-hBN interface is indirectly excited by the diffusion of free excitons generated in the volume of the hBN crystals towards the interface between them \cite{Roux2023}. The efficiency of the interface exciton luminescence, therefore, depends on two factors: the efficiency of free exciton transport from the bulk to the interface, and the efficiency of radiative recombination at the interface.

The luminescence from the twisted hBN-hBN homostructures is analyzed here as a function of the areal density of the excitation power (W/cm$^{2}$). Low power density data were recorded with an acceleration voltage of 5~kV with a constant current of 27~pA and spot diameters ranging from 1.6 to 26~µm. High power densities were obtained with a fixed spot diameter of 1.6~µm and excitation currents ranging from 27 to 5600~pA. This dual approach made it possible to vary the excitation power density over 5 orders of magnitude.
\begin{figure}[h!]
\includegraphics[scale=0.21]{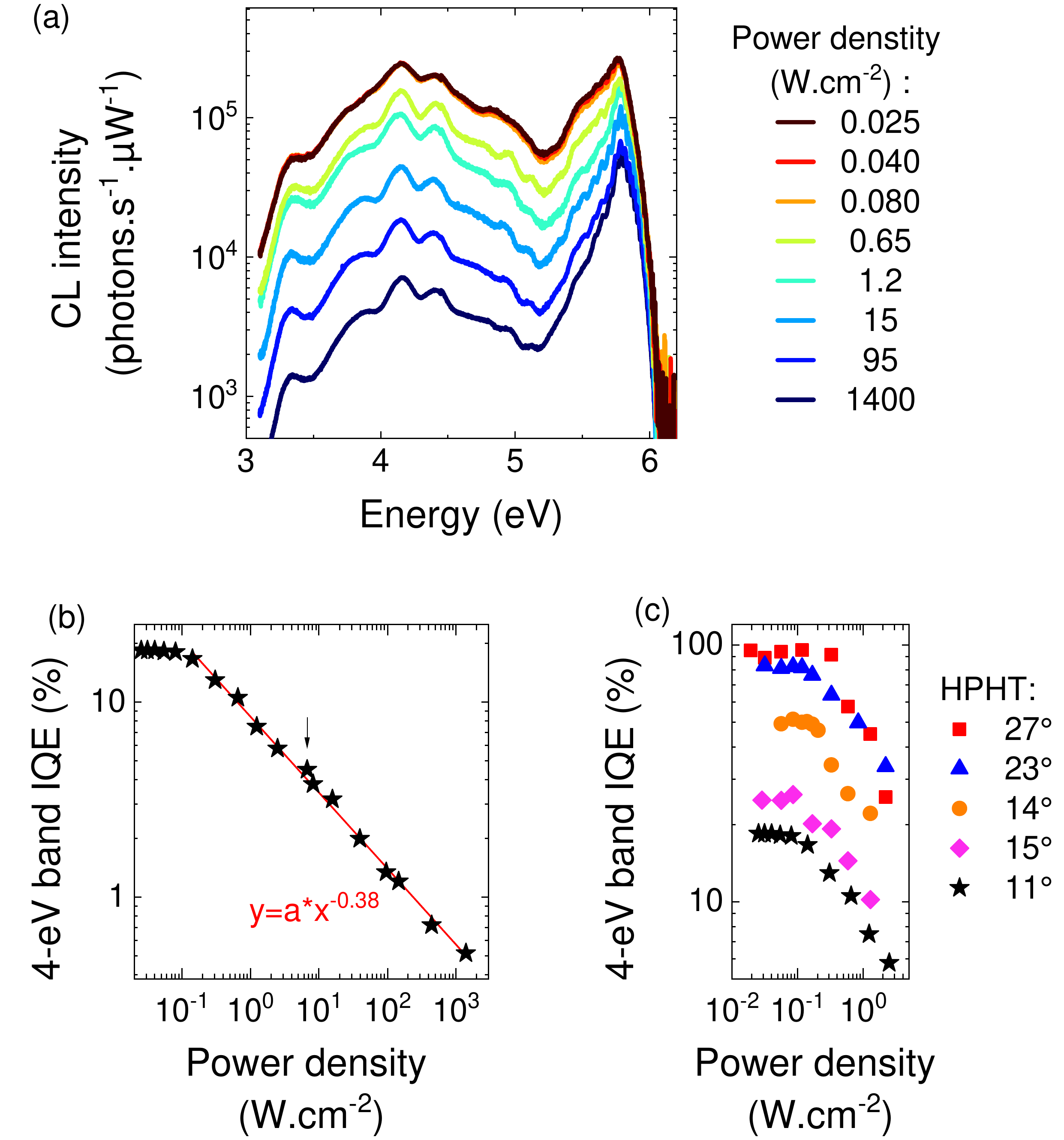}
\caption{(a) CL spectra normalized by the excitation power for different excitation densities (5kV, 300K) measured on the HPHT-11° sample (b) $IQE$ of the 4-eV emission, deduced from (a), as a function of the excitation power density. The black arrow indicates the transition point between the two methods for varying the excitation density, as indicated in the text. (c) $IQE$ of the 4-eV emission as a function of excitation density for the 5 HPHT samples. 
}
\label{F4}
\end{figure}

Figure~\ref{F4}(a) shows the CL spectra normalized by the excitation power for different excitation densities on the HPHT-11° sample. In Fig.~\ref{F4}(b), it is observed that above 0.1~W/cm$^{2}$, the efficiency of the 4-eV luminescence decreases significantly, with $IQE$ $\propto$ power$^{-0.38}$. This corresponds to a sublinear regime in CL intensity with I$_{CL}$ $\propto$ power$^{0.62}$, suggesting a bimolecular recombination of excitons at high power densities. Figure~\ref{F4}(c) indicates that this saturation effect appears in all HPHT homostructures with little effect of the twist angle on the 0.1~W/cm$^{2}$ saturation threshold.
In contrast, under similar excitation conditions, bulk hBN luminescence only saturates above 50~W/cm$^{2}$ due to exciton-exciton annihilation (EEA) of free excitons \cite{Plaud2019}\footnote{Experiments are performed at 10~K in ref. \cite{Plaud2019}, and at 300~K in ours. However, since in such samples the diffusivity is temperature independent \cite{Roux2023}, we do not expect drastic differences in the probability of exciton-exciton collisions and thus in the EEA rate as the temperature is increased.}. The much lower threshold in twisted hBN-hBN structures suggests that saturation is rather driven by the EEA of interface excitons. This interpretation is consistent with the much higher exciton density at the interface than in the volume, which results from the efficient capture of excitons at the interface, as shown later in this section.

Interestingly, at low power densities (below 0.1~W/cm$^{2}$), the $IQE$ of the 4-eV luminescence can reach very high values, as overviewed in Tab.~\ref{T2}. We observe that the $IQE$ increases with the twist angle, reaching nearly 100\% at a 30° twist. This suggests that, for 30° twist angles, the vast majority of excitons are trapped at the interface and recombine via photon emission in the 4-eV band. Such a highly efficient emission is rare at 300~K and may be promising for light emitting devices. A unity $IQE$ of the interface luminescence implies that (i) the transport of free excitons from the bulk to the interface is without losses, and (ii) that the exciton recombinations at the interface are fully radiative. Previous results on exciton diffusion in bulk hBN monocrystals suggest that the conversion of 3D excitons into 2D interface/surface excitons is extremely efficient even for dark interfaces (with low twist angles) and for the non-radiative free surface \cite{Roux2023}. This study further shows that the recombination of the 2D interface excitons is mostly radiative near the 30° twist angle quasi-crystal limit.

\begin{table}[h]	
\begin{tabular}{|c|c|}
\hline
Sample  &4-eV $IQE$ (\%) \\
\hline
HPHT-11° & 20\\
HPHT-14° & 50\\
HPHT-15° & 30\\
HPHT-23° & 80\\
HPHT-27° & 90\\
\hline
\end{tabular}
\caption{$IQE$ of the 4-eV emission measured below 0.1~W/cm$^{2}$ on twisted hBN homostructures made of HPHT crystals.}
\label{T2}			
		\end{table}%

Note that between 10$^{-2}$ and 10$^{-1}$ W/cm$^{2}$, the average energy of the photons emitted within the 4-eV band is slightly red-shifted by about 15~meV as the excitation power is increased. In contrast, a hypothetical donor-acceptor pair (DAP) recombination, which would have been a candidate for such a broad emission, should exhibit a blue-shift when increasing the power excitation \cite{Peter2010, Pankove1975}. Our observations then discard a DAP origin for the 4-eV emission. 

As a summary of this section, power-dependent experiments reveal a strong decrease in the efficiency of the 4-eV emission at moderate excitation, probably caused by a bimolecular annihilation process between excitons accumulated at the hBN-hBN interface. At low excitation, in the most favorable cases, i.e., at twist angles close to 30° with HPHT crystals presenting a long-range exciton diffusion, most excitons recombine radiatively at the interface with a 4-eV photon emission.

\subsection{V– Exciton self-trapping at the twisted hBN-hBN interface}

The characteristics of the 4-eV emission (2-eV energy shift and the 2-eV broadening) supports a self-trapping mechanism of excitons at the interface that we develop in this section. 

\begin{figure}[h!]
\includegraphics[scale=0.35]{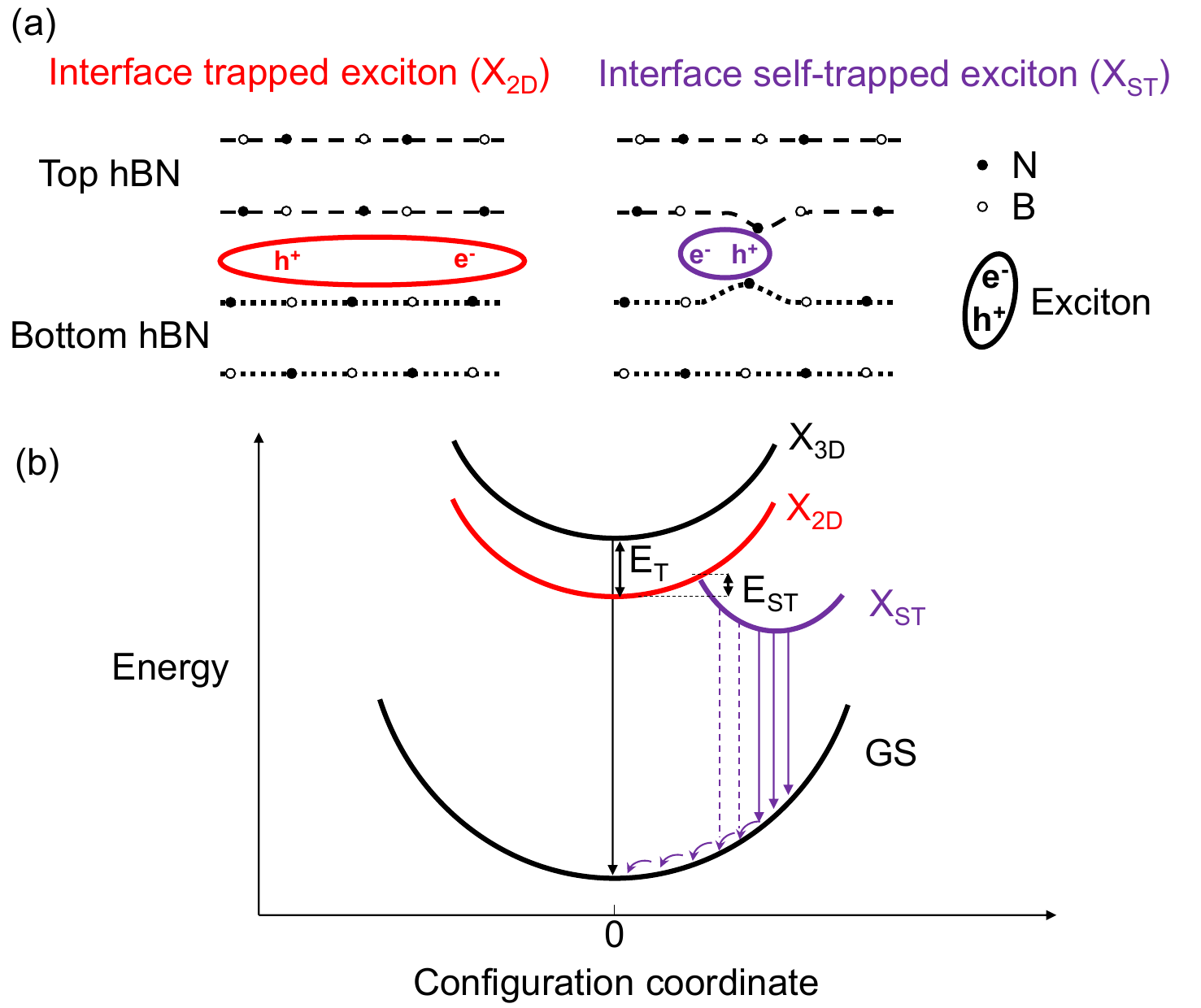}
\caption{(a) Schematics of a trapped exciton at the interface (X$_{2D}$), in red, and of a self-trapped exciton at the interface (X$_{ST}$), in purple, with a self-induced lattice deformation on which the exciton is trapped. (b) Configuration diagram representing the three excitons populations in hBN-hBN structures: X$_{3D}$, X$_{2D}$ and X$_{ST}$. The exciton trapping potential of the interface at zero distortion appears as $E_{T}$ between X$_{3D}$ and X$_{2D}$. The energy barrier for the formation of the self-trapped exciton is indicated ($E_{ST}$). The long straight arrows represent the photons resulting from the radiative recombination of free excitons (in black), of thermalized self-trapped excitons (purple, solid lines), and of non-thermalized self-trapped excitons or "hot" self-trapped excitons (purple, dashed line). The small curved arrows represent the phonons required for the lattice to return to the ground state during the recombination of a self-trapped exciton.
}
\label{F5}
\end{figure}

Figure~\ref{F5}(a) illustrates the difference between simple exciton trapping in the potential well formed by the interface between the two twisted hBN flakes (X$_{2D}$) and self-trapping of this exciton at the interface (X$_{ST}$). A self-trapped exciton is an exciton that induces a lattice distortion around it, and that is trapped in this distorted area. 

The configuration diagram drawn in Fig.~\ref{F5}(b) displays the exciton energy as a function of the lattice deformation around it, represented by the configuration coordinate. In this diagram, we define the trapping energy potential of the twisted interface at zero deformation, denoted $E_{T}$. Its value for the twisted interface in hBN is a priori estimated to be in the order of 100~meV following recent theoretical calculations \cite{Latil2023, Su2022}. 

The energy barrier for the self-trapping phenomenom is denoted $E_{ST}$ in Fig.~\ref{F5}(b). Its existence was predicted by Landau \cite{Landau1933} and Toyozawa \cite{Toyozawa1958} for electrons and by Rashba \cite{Rashba1957, Rashba1982} and Sumi and Toyozawa \cite{Sumi1971} for excitons. The presence of an energy barrier for self-trapping was evidenced experimentally in the 60's and 70's in alkali halides \cite{Blume1969, Kink1969, Lushchik1970}. The energy barrier disappears only for self-trapping of a localised exciton  on a 0D site \cite{Rashba1982, Williams1990}, while here the hBN-hBN interface is a 2D defect. 

The presence of the energy barrier $E_{ST}$ implies the coexistence of excitons simply trapped at the 2D interface (X$_{2D}$) and excitons self-trapped at the interface (X$_{ST}$). Considering the free excitons in the 3D bulk hBN (X$_{3D}$), there are thus three coupled exciton populations during the CL experiments: X$_{3D}$, X$_{2D}$ and X$_{ST}$.

Note that, though $E_{ST} \neq 0$, the energy barrier for self-trapping is necessarily low, unless an efficient tunnelling occurs, since luminescence of X$_{ST}$ is still observed close to liquid helium temperature (see Appendix E). However, its presence basically limits the formation of X$_{ST}$ and reduce the intensity of their luminescence as the temperature is lowered \cite{Blume1969, Kink1969, Lushchik1970}.

The general features of self-trapped excitons are well consistent with our observations. Indeed, the radiative recombination of a self-trapped exciton is accompanied by the emission of a set of phonons to restore the crystal lattice to its undistorted form as sketched in Fig.~\ref{F5}(b). This drastically reduces the energy of the emitted photon with respect to the band gap of the interface and explains the large energy shift ($\approx$2~eV) between the excitonic bandgap and the luminescence at the interface. Furthermore, the luminescence of self-trapped excitons is inherently broad for several reasons. First, due to local disorder at the interface (which is likely to be high at large twist angles, see Fig. 1(d)), the variation of the local potential landscape might cause self-trapped excitons to exhibit a variety of energies and distortions. Second, the recombination of self-trapped excitons requires multiple phonon emissions, which further increase the broadening of their luminescence. This is illustrated in Fig.~\ref{F5}(b), where the three purple arrows (solid lines) represent photons of different energies resulting from the recombination of different self-trapped excitons at the band edge.

Experimentally in this study, the emission related to excitons simply trapped at the interface (X$_{2D}$) has not been detected in CL. Two possible scenarios can be considered to account for an almost zero-luminescence intensity: either 1) the X$_{2D}$ population is empty, or 2) the X$_{2D}$ excitons are dark, i.e. have an infinite radiative lifetime. In the first scenario, excitons would accumulate at the interface in the self-trapped form as a X$_{ST}$ state. This requires their self-trapping to be almost instantaneous, which contradicts the presence of an energy barrier to self-trapping limiting the formation of X$_{ST}$. Scenario 1) also implies that the detrapping of excitons from the interface into the volume occurs from the X$_{ST}$ states, which would be energetically too costly. The first scenario clearly appears inconsistent with the significant release of free excitons in the bulk hBN volume from the interface observed by TRCL. Therefore, we have ruled out scenario 1) and in the following we will only consider scenario 2).

In this scenario, X$_{2D}$ excitons have a negligible probability of radiative recombination (dark excitons). This is in good qualitative agreement with (i) recent investigations \cite{Roux2023} which show that the free surface of hBN single crystals acts as a non radiative trap for excitons, (ii) the long lasting release of the interface excitons in the volume, and (iii) the $IQE$ close to 100\% observed for X$_{ST}$ at high twist angles (Tab.~\ref{T2}), implying that the recombination rate of X$_{2D}$ is very low compared to that of X$_{ST}$. Since the IQE reaches almost unity at high angles (see table~\ref{T2}.), the localised self-trapped excitons then recombine very fast and radiatively, with an almost instantaneous radiative recombination rate compared to the other processes that have been considered. For this reason, we will further neglect the eventual non-radiative recombinations of X$_{2D}$. Figure \ref{F7} further summarizes these conclusions.

\subsection{VI - Temperature-dependent experiments}

To strengthen the attribution of the 4-eV emission to a self-trapping process, and to characterize the different exciton populations in more detail, we have performed a series of experiments as a function of temperature. 

\begin{figure}[h!]
\includegraphics[scale=0.21]{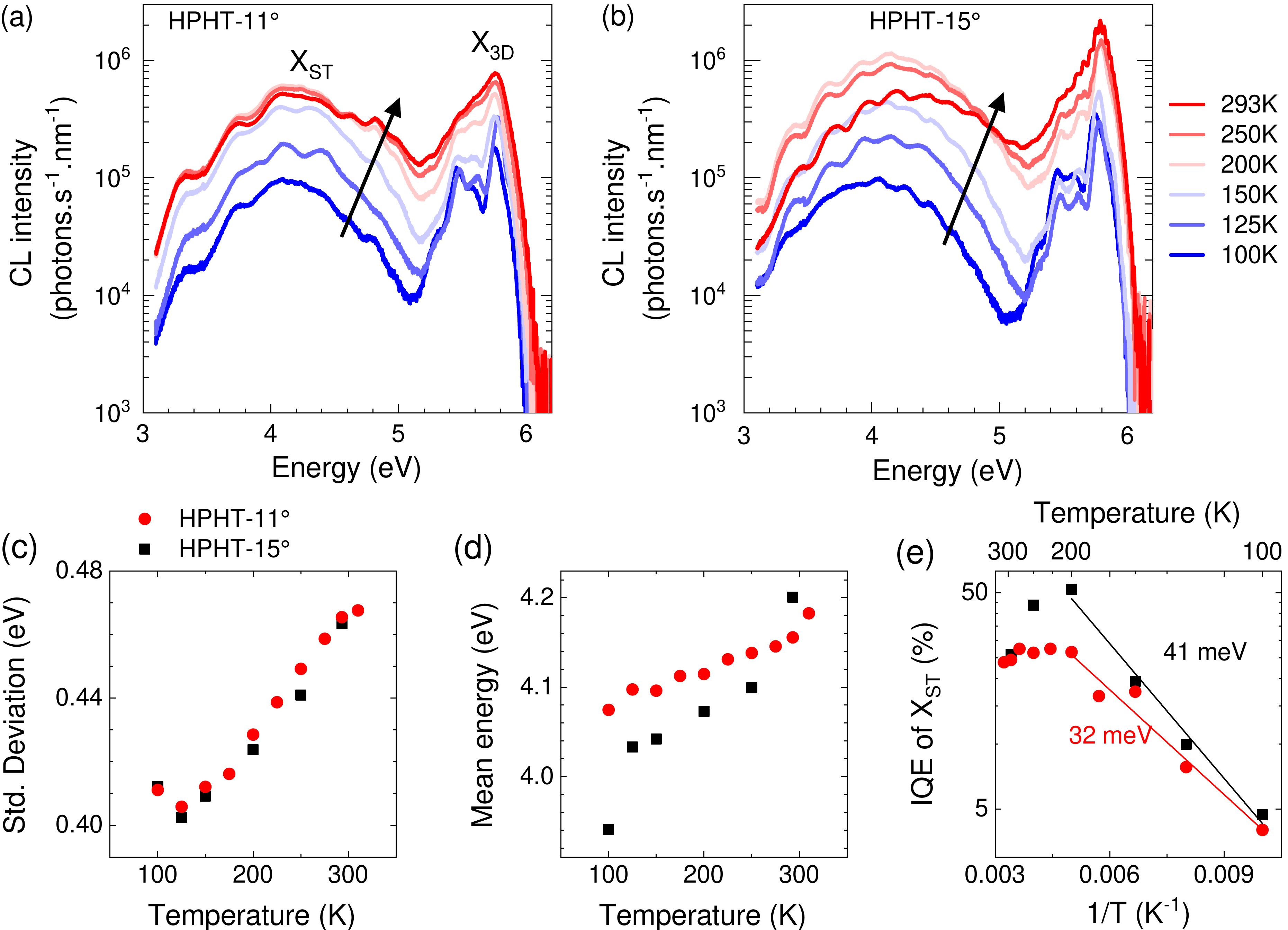}
\caption{CL spectra recorded on the HPHT-11° (a) and HPHT-15° (b) samples between 100 and 293~K. (c) Standard deviation of the 4-eV emission energy, (d) average emission energy, and (e) $IQE$ of the self-trapped exciton X$_{ST}$ luminescence as a function of temperature, extracted from the spectra in (a-b). Arrhenius laws are shown in full lines in (e), from which activation energies are extracted.
}
\label{F6}
\end{figure}

The intermediate twist angle samples HPHT-11° and HPHT-15° are studied at a very low power density (5~kV, 27~pA, 60~µm spot diameter; $\approx$ 0.005~W/cm$^{2}$), which prevents saturation effects down to 100~K. Since saturation effects appears to be stronger at low temperature and could not be be avoided below 100~K, we limited the data analysis to the 100-300~K range. We first studied the influence of temperature on the spectral features of the 4-eV luminescence with Fig.~\ref{F6}(a-b).

Non-thermalized (hot) self-trapped excitons are known to be favored at high temperatures, where their recombination, shown in Fig.~\ref{F5}(b), occurs at a higher energy than when thermalized. We therefore expect an increase in the emission energy of self-trapped excitons as the temperature increases \cite{Lushchik1982, Tekhver1974}. Between 200 and 300~K, Fig.~\ref{F6}(a-b) indeed depicts an increase of the X$_{ST}$ exciton luminescence in the 4-5~eV region with respect to the 3-4~eV region, consistent with a contribution from non-thermalized X$_{ST}$. The broadening of the 4-eV emission on the high-energy side, indicated by black arrows in Fig.~\ref{F6}(a-b), is quantitatively characterized by the standard deviation of the emission energy (Fig.~\ref{F6}(c)) and by the average energy of the X$_{ST}$ emission (Fig.~\ref{F6}(d)). Both slightly increase at high temperature. Note that the spectral width of the hBN-hBN 4-eV luminescence still remains around 2~eV at cryogenic temperatures, as shown in the Appendix E. This rules out the possibility of color center emission which would exhibit narrow linewidths at 5K \cite{Vuong2016, Fournier2021}.

Figure~\ref{F6}(e) shows that the $IQE$ of the X$_{ST}$ luminescence drops below 200~K, and is only a few percent at 100~K. The activation of X$_{ST}$ emission with temperature is consistent with the presence of an energy barrier for self-trapping as discussed in the previous section (named $E_{ST}$ in Fig.~\ref{F5}(b)). With an Arrhenius plot fit, we find $E_{ST}$ values of 32 and 41~meV for HPHT-11° and HPHT-15° samples respectively. These values are probably overestimated due to the saturation effects evidenced in Section IV, which are not completely avoided near 100~K. They are also subject to a high uncertainty due to the small temperature range investigated. It is therefore reasonable to estimate the activation energy of the self-trapping to be a few tens of meV in these samples. This corresponds well to the order of magnitude found in the literature for the self-trapping of excitons in other materials such as alkali halides \cite{Lushchik1982, Jung2021, Blume1969, Lushchik1970, Kobitski2001}.

\onecolumngrid

\newpage
\twocolumngrid

As a summary, the various effects revealed by the temperature-dependent experiments appear to be well consistent with a 4-eV emission band resulting from the recombination of excitons that are self-trapped at the twisted hBN-hBN interface.

\subsection{VII - Phenomenological model for exciton recombination dynamics}

\begin{figure}[h!]
\includegraphics[scale=0.5]{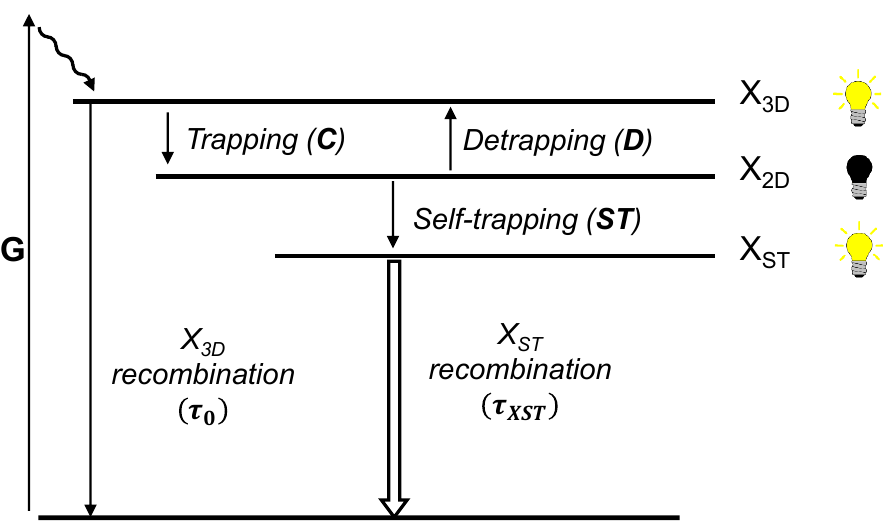}
\caption{Energy levels and transitions between the three exciton populations within the twisted hBN-hBN homostructures. In this scenario, excitons from volume (X$_{3D}$) are trapped at the interface, where they accumulate in the dark and long living form (X$_{2D}$). Once self-trapped (X$_{ST}$), they recombine instantaneously with a photon emission, leading to the intense light emission around 4~eV.}
\label{F7}
\end{figure}

We now explain the recombination dynamics of the three exciton populations in hBN-hBN homostructures (X$_{3D}$, X$_{2D}$ and X$_{ST}$), considering the interactions that bind them, in order to ascribe a physical meaning to the decay times measured in TRCL. Different transitions between these exciton populations are shown in Fig.~\ref{F7} as a summary of section V conclusions. X$_{2D}$ accumulate at the interface after their capture at a rate $C$ from free excitons of the hBN bulk crystals generated by the primary excitation. They are then either released to the volume as X$_{3D}$, with a detrapping rate $D$,  or self-trapped in the X$_{ST}$ states with a rate $ST$.

\begin{figure}[h!]
\includegraphics[scale=0.19]{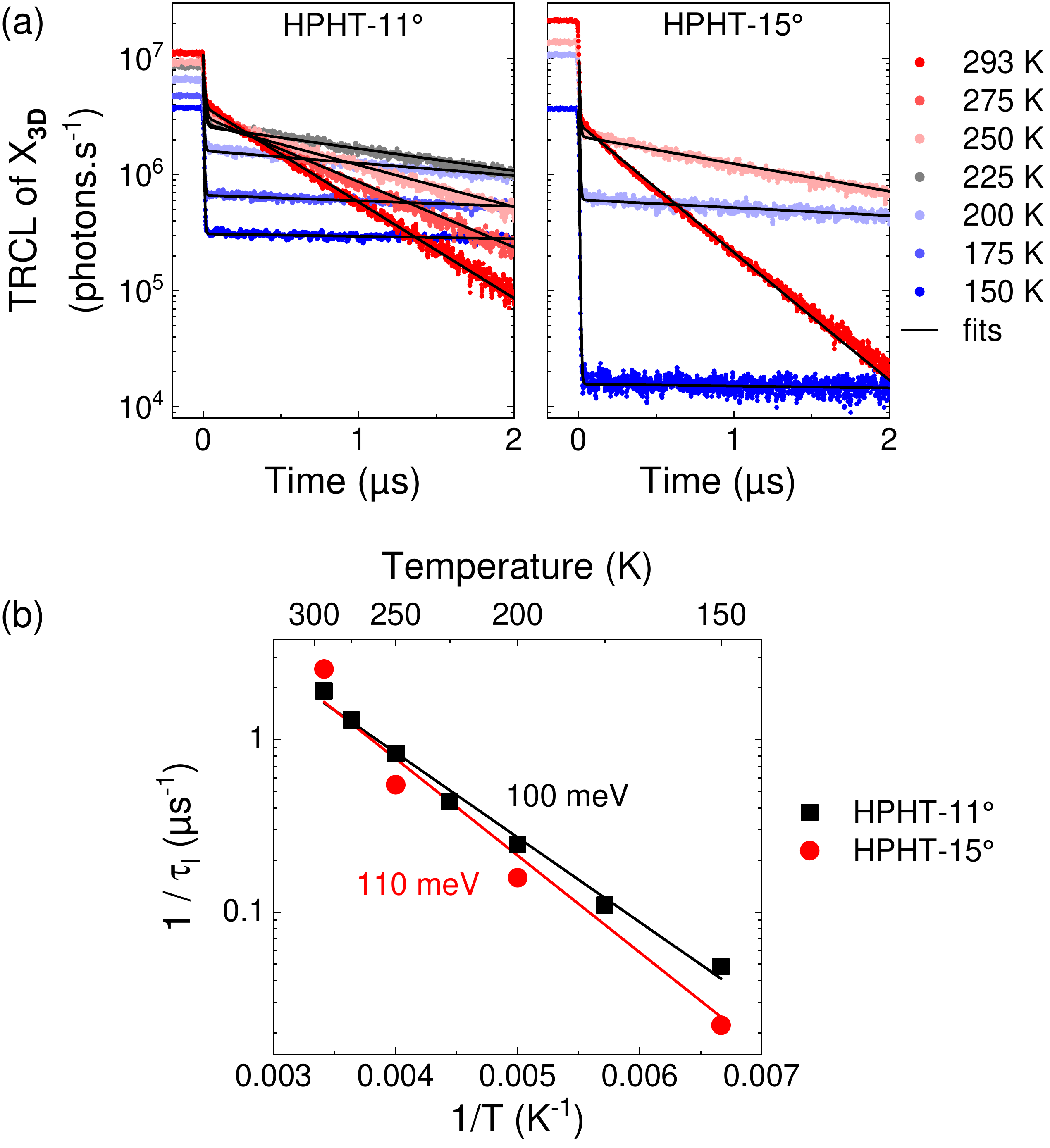}
\caption{(a) TRCL decays of the free exciton luminescence for temperatures in the range 150 - 293~K on HPHT-11° and HPHT-15 samples. Bi-exponential decay fits are used to extract the characteristic time of the long component $\tau_{l}$. (b) 1/$\tau_{l}$ as a function of 1/T in HPHT-11° and HPHT-15° samples. Arrhenius laws appear as solid lines.}
\label{F8}
\end{figure}

According to this scenario, the long decay time of X$_{3D}$ and X$_{ST}$ measured in TRCL ($\tau_{l}$) after the interruption of the continuous excitation (see Fig.~\ref{F3}) corresponds to the lifetime of the X$_{2D}$ excitons decaying either by detrapping or self-trapping. Considering the associated rates, $\tau_{l}$ is given by: 1/$\tau_{l} \approx D+ST$ (details given in the Appendix F).
In the following we discuss the temperature-dependence of the long decay time common to X$_{3D}$ and X$_{ST}$. Since these decay times are found equal for X$_{3D}$ and X$_{ST}$ populations when decreasing the temperature, we present only the TRCL measurements of X$_{3D}$. Fig.~\ref{F8}(a) shows the decays for two samples with intermediate twist angles: HPHT-11° and HPHT-15°. The decays are normalized by the absolute integrated intensity of the X$_{3D}$ band obtained by independant CL measurements under continuous excitation (spectra in Fig.~\ref{F6}(a-b)). The characteristic time of the long component $\tau_{l}$ is extracted from bi-exponential function fits.

Figure~\ref{F8}(b) shows a plot of 1/$\tau_{l}$ $\approx$ $D$+$ST$ as a function of 1/T(K). The data are fitted with Arrhenius functions, yielding activation energies of 100 and 110~meV for HPHT-11° and HPHT-15° samples, respectively. Given the low activation energy estimated for the formation of self-trapped excitons ($E_{ST} \simeq $ 10 meV), 1/$\tau_{l}$ $\approx$ $D$+$ST$ is likely limited instead by the thermal activation of the detrapping rate $D$, with an activation energy corresponding to $E_{T}$. The self-trapping of an exciton requires the assistance of many phonons for the distortion to occur. Despite a relatively low activation energy, the self-trapping may be extremely slow at room temperature, in good agreement with next results in this section. It indicates that $D \gg ST$ between 150 and 300~K, consistently with the relatively low efficiency of the 4-eV luminescence observed in these samples at 300~K (see Tab.~\ref{T2}). We conclude that the extracted activation energies correspond to the depth of the potential well experienced by bulk excitons at the twisted hBN interface, E$_{T}$, which is found to be around 100~meV for twist angles between 10 and 15°. This value is in good agreement with the first theoretical estimates of E$_{T}$\cite{Latil2023, Su2022}.

Coming back to the high twist angles, we remind that the efficiency of the 4-eV emission is close to 100\% near 30° twist angles at 300~K (see Tab.~\ref{T2}). This requires an almost full conversion of $X_{2D}$ into $X_{ST}$ states via self-trapping. To achieve this, the detrapping rate is necessarily negligible compared to the self-trapping rate $D \ll ST$ , and the lifetime of the interface excitons is then limited by the self-trapping 1/$\tau_{l}\approx ST$. The situation is opposite to the 10-15° twisted hBN crystals, where 1/$\tau_{l}\approx D$. This comparaison highlights that the interpretation of the long decay time observed in TRCL is cautious depending on twist angles.
\begin{figure}[h!]
\includegraphics[scale=0.16]{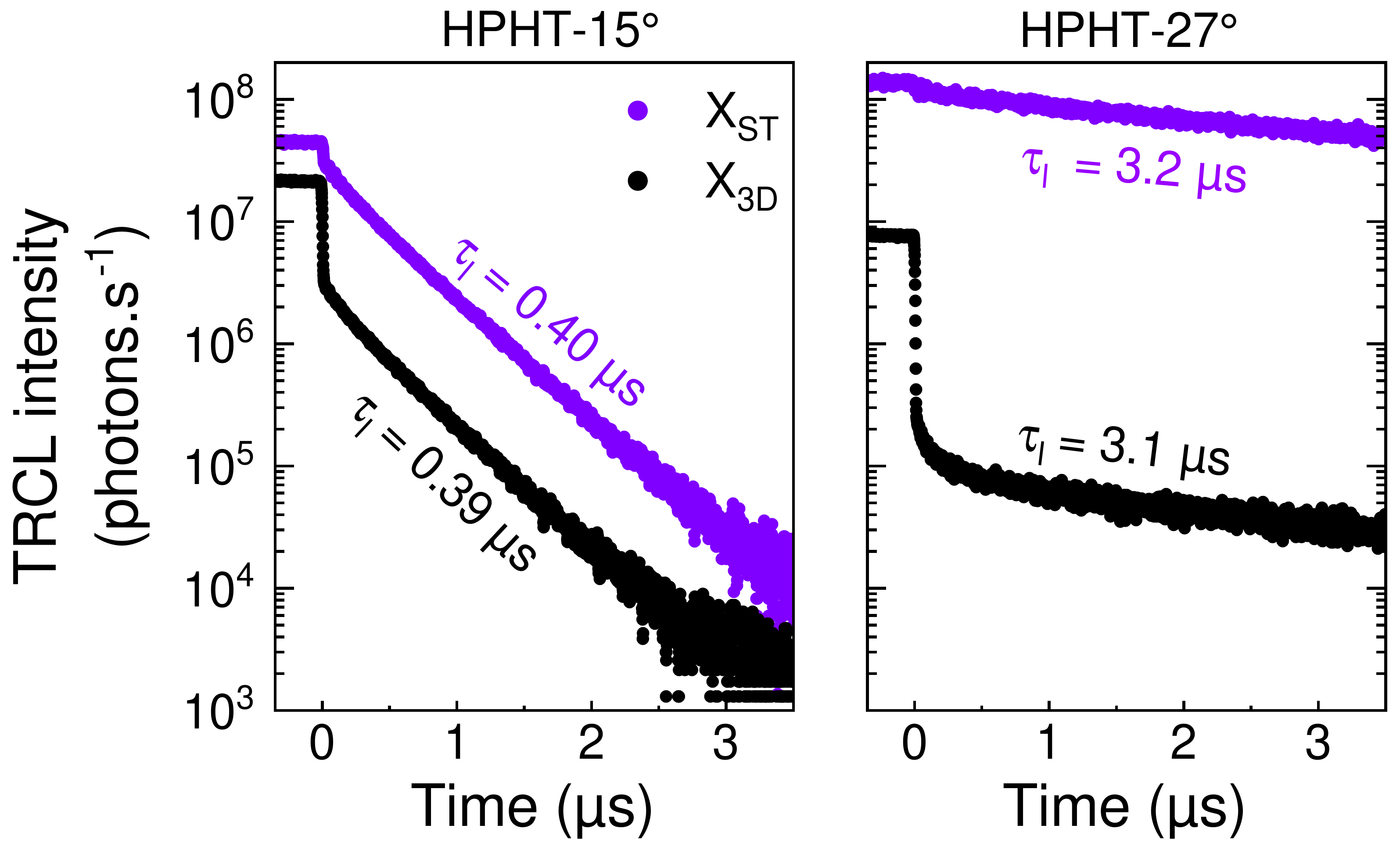}
\caption{(a) TRCL decays of the free exciton luminescence (in black) as compared to the self-trapped exciton luminescence (in purple) for an intermediate twist angle sample HPHT-15° and a high twist angle sample HPHT-27°. Experiments done at 300~K.}
\label{F8bis}
\end{figure}

We further analyse and compare the TRCL results between 15° and 27° twists at 300~K in Fig.~\ref{F8bis}. At a 15° twist angle, we then attribute the long decay time to the detrapping rate $D$, found equal to 2.5*10$^{6}$s$^{-1}$ at 300 K while the self-trapping rate $ST$ is much lower but unknown. At 27° twist angle, we access the self-trapping rate $ST = $2.5*10$^{5}$ s$^{-1}$, while the detrapping rate $D$ is much lower but unknown. These limit cases are very interesting compared to intermediate twist angles where both should contribute to the long decay time with 1/$\tau_{l}$ $\approx$ $D$+$ST$.

The self-trapping process is found to be particularly slow around 30° twist (3.2 µs on HPHT-27°, 4 µs on HPHT-24°) though giving high efficiency of the 4-eV emission. Around 15° twist, the dominant detrapping is about 10 times faster (0.40 µs on HPHT-15°, and 0.46 µs on HPHT-11°) with a typical efficiency decrease of the 4 eV emission by a factor of 2 (see Tab.~\ref{T2}). This suggests that the detrapping of excitons from the interface is decreasing at high angle, giving enough time for exciton trapped at the interface to slowly deform the crystal locally around it for self-trapping. This means that the depth of the potential well formed by the twisted hBN interface, E$_{T}$, found equal to 100 meV at 15°, increases near 30° so that detrapping becomes negligible in front of self-trapping. Theoretical calculations of E$_{T}$ for twist angles between 15 and 30° would be welcome to confirm that point.

In summary, we interpret our experimental data with excitons accumulating at the interface in a non-radiative and long-lived form (X$_{2D}$), followed by their subsequent conversion into $X_{3D}$ and $X_{ST}$ via their detrapping or self-trapping. Our results suggest that the depth of the interface trapping potential for free excitons in hBN increases with the twist angle. For low angles the detrapping is believed to be dominant and limits the efficiency of the interface luminescence, while close to 30°, the detrapping is negligible and the efficiency of the interface emission reaches almost 100\%. Our phenomenological model provides a good description of full the experimental data set, including the long-lasting luminescence and the high $IQE$.

\subsection{VIII - Discussion: Self-trapping in sp$^{2}$ BN}
We finally discuss the origin of the exciton self-trapping in hBN. Exciton self-trapping is favored by the ability of the lattice to deform around the exciton. Given the high ionicity of the covalent B-N bonds, a strong exciton-lattice interaction is expected in hBN. In the excitons of bulk hBN, the electron and the hole distance are distant from $\approx$7~$\AA$ \cite{Roux2021}, which makes the hBN excitons particularly small. Still an exciton dipole of this size is probably too large to interact with the atomic bonds in bulk hBN and to give rise to self-trapped excitons.

The exciton behaviour at hBN/hBN interface still needs further investigations, but we can already draw general lines. When an exciton is spatially confined, it is a general trend that its binding energy increases and its size decreases. Theoretical calculations have been reported for the free surface of hBN \cite{Paleari2018} and for the BN monolayer. In the latter, the exciton size is calculated at $\approx$ 2~$\AA$ \cite{Galvani2016}. A similar trend is expected for 2D excitons confined at the hBN/hBN interface. We then suggest that the exciton size decreases when confined at the interface with respect to the bulk, sufficiently to match the order of the lattice parameter (2.5 $\AA$ in plane, 3.3 $\AA$ out of plane), then promoting the exciton-lattice coupling at the interface by dipole-dipole interactions.

For small twist angles of a few degrees, the interfaces are known to be energetically stabilized by Moiré formation and  are probably weakly deformable after that. Indeed, at low angles, intense atomic reconstruction occurs and mechanically stabilizes the structure by expanding the region with stable stacking configurations and condensing all the strain and the unstable stacking configurations into 1D lines \cite{Yasuda2021, Woods2021}. On the contrary, near 30° of twist where self-trapping is better observed, a transition from a commensurable to an incommensurable interface occurs, as studied in quasicrystals, leading to a drastic change in the structure and the physical phenomena it can host. While at low angles the perturbation of the charge distribution induced by the appearance of the $\pi^{B-B}$ and $\pi^{N-N}$ homo-bonds is weak and confined to certain zones of the Moiré structure, at high angles a very large fraction of the $\pi$ bonds are perturbed and densely distributed throughout the interface. As a consequence, the potential perturbations increase in locality and density with increasing twist angle, which could promote a higher barrier for exciton detrapping and favor the slow self-trapping process.

These elements address the role of the twist angle on exciton interactions with the interface, but also highlight the need for theoretical studies to better understand the origin of exciton self-trapping by exploring (i)  what kind of hBN lattice distortions are involving self-trapped excitons  and (ii) how the potential well seen by hBN excitons does evolve between 5 and 30° twisted interface.

This work on self-trapped excitons at twisted hBN-hBN interfaces might cast a new light on similar luminescence signal reported in defective hBN samples \cite{Museur2008} as well as in other sp$^{2}$-hydridized BN samples that exhibit interface disorder, such as turbostratic BN, pyrolytic BN, or BN multiwall nanotubes \cite{Museur2009, Jaffrennou2008}. Also typically 2~eV broad, centered around 4 eV and with long decay dynamics, this emission is excited only above 6~eV excitation, i.e. above the excitonic bandgap, suggesting an excitonic origin. This similar emission observed in disordered BN materials could also be related to exciton self-trapping, which would then appear to be a fairly common phenomenon in the BN material familly.

Recent studies report unusual exciton-lattice interactions at other h2D interfaces \cite{Merkl2021, Chow2017, Hennighausen2023, Deng2022}, which in some cases could lead to the formation of self-trapped excitons \cite{Deng2022}. Our results added to these studies show that the properties of these self-trapped states are easily modulated by the structural properties of h2Ds, such as the twist angle, the choice of stacking, or the thickness of the crystals. h2Ds appear to be an ideal platform for studying and manipulating self-trapped states, opening up new fields of application for these materials. In particular, the high efficiency and the large broadening of the 4-eV luminescence of self-trapped excitons at twisted hBN-hBN interfaces could be exploited for broadband UV light sources \cite{Wang2018, Jun2019, Li2022b, Lian2021}.

Finally, the formation of interlayer excitons at the interfaces between two TMDs, (with hole and electron spatially separated in the two layers), has been shown to increase the exciton lifetime by more than 3 orders of magnitude \cite{Rivera2015, Ovesen2019, Miller2017, Yuan2020}. The long lifetime and dark character of the X$_{2D}$ excitons at the hBN-hBN interface might have common features with  interlayer excitons.

\subsection{IX - Conclusion}
Sixteen twisted hBN-hBN homostructures have been studied using CL and TRCL as function of temperature with the aim of understanding the broad 4-eV luminescence in hBN-hBN homostructures. Our study shows that this luminescence is indirectly excited by the transport of free excitons from the upper and lower hBN flakes to the interface between them via their diffusion and capture. Accumulated at the interface, excitons probably undergo exciton-exciton annihilation which manifests by the decrease of the light emission efficiency at moderate excitation powers. In the low excitation regime, it was found that the efficiency of the 4-eV emission reaches almost 100\% at angles close to 30°. This intense light emission is attributed to the radiative recombination of self-trapped excitons at the interface. A reduction of the excitons size after their capture at the interface is believed to favor a local deformation of the crystal around them in which they self-trap.

TRCL luminescence decays could be analyzed by considering a trapping-detrapping phenomenological model where 3 exciton populations are coupled: free excitons in the volume (X$_{3D}$), excitons trapped at the 2D interface (X$_{2D}$, not self-trapped), and excitons self-trapped at the interface (X$_{ST}$). In this scenario consistent with our set of experimental data, excitons accumulate at the interface in the non-radiative (dark) and long living X$_{2D}$ form. Our analysis of temperature-dependant experiments for twist angles of $\sim$15° reveals a self-trapping energy barrier of a few 10 meV and an interface trapping potential measured around 100~meV. The quantitative analysis of TRCL data further provides an estimation of the detrapping rate near 15° and of the self-trapping rate near 30°. They suggest a deeper interface potential near 30° twist angles, limiting exciton detrapping from the interface and thus favoring self-trapping. This is consistent with the remarkably high luminescence efficiency of the X$_{ST}$ luminescence at high twist angles. These values agree with the literature and theoretical calculations available to date.

For highly twisted hBN-hBN structures, the crystal lattice appears to be locally and densely perturbed across the entire interface. Beyond these qualitative elements, it remains to be understood in detail what is the key elements that cause self-trapping. The physical phenomena occurring in h2Ds at high angles appear very different from those at small angles with moiré superstructures. To explore them, one avenue would be to apply the quasi-crystal physics to h2Ds with high twist angles.

\subsection*{Acknowledgments}
\begin{acknowledgments}
The research leading to these results has received funding European Union's Horizon 2020 research and innovation program under grant agreements No 785219 (Graphene Core 2) and No 881603 (Graphene Core 3). Support for the APHT hBN crystal growth comes from the Office of Naval Research, Award No. N00014-20-1-2474. K.W. and T.T. acknowledge support from the JSPS KAKENHI (Grant Numbers  21H05233 and 23H02052) and World Premier International Research Center Initiative (WPI), MEXT, Japan. This work was also supported by Agence Nationale de la Recherche funding under the program ESR/EquipEx+ (grant number ANR-21-ESRE- 0025) and ANR ATOEMS .
\end{acknowledgments} 
\subsection*{Appendices}
\subsection*{Appendix A: Measurement of the internal quantum efficiency of the 4-eV luminescence.}
The procedure to measure the internal quantum efficiency ($IQE$) of a luminescence band during CL experiment under continuous excitation is described in ref. \cite{Schue2019}. In this reference, it is applied to measure the $IQE$ of the free exciton luminescence of hBN single crystals that appears at 5.75~eV. To apply the same procedure to the 4-eV band, one should consider the significative change of the refractive index of the top surface between 5.75~and 4-eV. 

\begin{figure}[h!]
\includegraphics[scale=0.23]{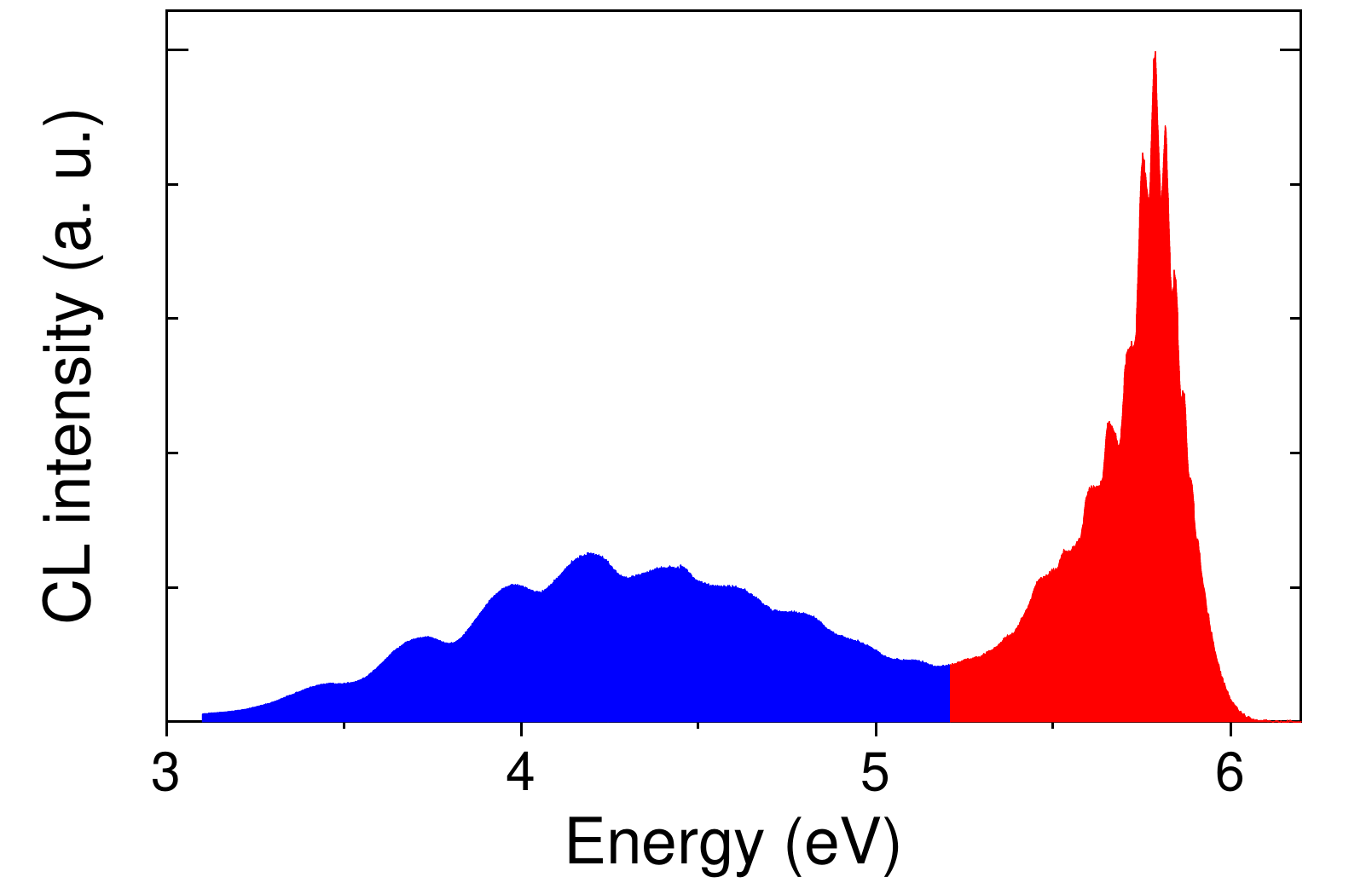}
\caption{Corrected CL spectrum of a hBN-hBN homostructure at 300~K. The integrated intensity of the free exciton band of hBN appears in red, and the integrated intensity of the 4-eV luminescence, characteristic of twisted hBN-hBN homostructures, appears in blue.}
\label{F9}
\end{figure}

In our case, the CL intensity is integrated between 3.1 and 5.2~eV as shown in Fig.~\ref{F9}. The reflection index $n$ of the surface is increased from $n$=2.37 to $n$=3 over the 3.1-5.2~eV range \cite{Segura2018}, corresponding to a light extraction that varies from 3.9\% to 2.1\%. To simply measure the magnitude of the $IQE$ of this band, we approximate a constant index $n$=2.56 measured at the band maximum (3.2\% extraction). This approximation is reasonable given the uncertainty of the $IQE$ measurement, which is of the order of 50\% \cite{Schue2019}.
\subsection*{Appendix B: Measurement of the twist angle by electron diffraction.}
The twist angle of hBN-hBN homostructure was determined by Electron Backscatter Diffraction (EBSD) and Electron Channeling Pattern (ECP).

In EBSD, the surface crystal orientations of a sample in the SEM are determined for each diffraction pattern using the OIM Analysis software from EDAX. For hBN-hBN homostructures, the crystal orientation of the top crystal is expressed in the reference frame of the bottom crystal. Not surprisingly, the out-of-plane z-axis of the two hBN crystals is identical. On the other hand, different in-plane orientations are measured as shown in Fig.~\ref{F1}(b) in the main text. The twist angle between the two crystals corresponds to this difference in orientation. It is equivalent to the smallest angle that can be found between the atomic B-N bonds of the two crystal lattices, regardless of their orientation (B to N or N to B). Between two hBN crystals this angle varies between 0 and 30°. EBSD mapping can quickly measure the orientation of several crystals separated by a few millimeters with an accuracy of the order of 1° (Fig.~\ref{F1}(b)).

In ECP, mapping is not possible, but the diffraction pattern on a selected zone of a 2D crystal gives its in-plane crystal orientation directly, without the need for complex processing. By comparing the ECP images of two 2D crystals placed on top of each other, we can measure the twist angle between the two crystals with an accuracy of the order of 1°. Its absolute value varies between 0 and 30° and corresponds to the angle measured by EBSD, as shown in Fig.~\ref{F1}(c).

\subsection*{Appendix C: CL signal on low angle hBN-hBN structures.}
\begin{figure}[h!]
\includegraphics[scale=0.23]{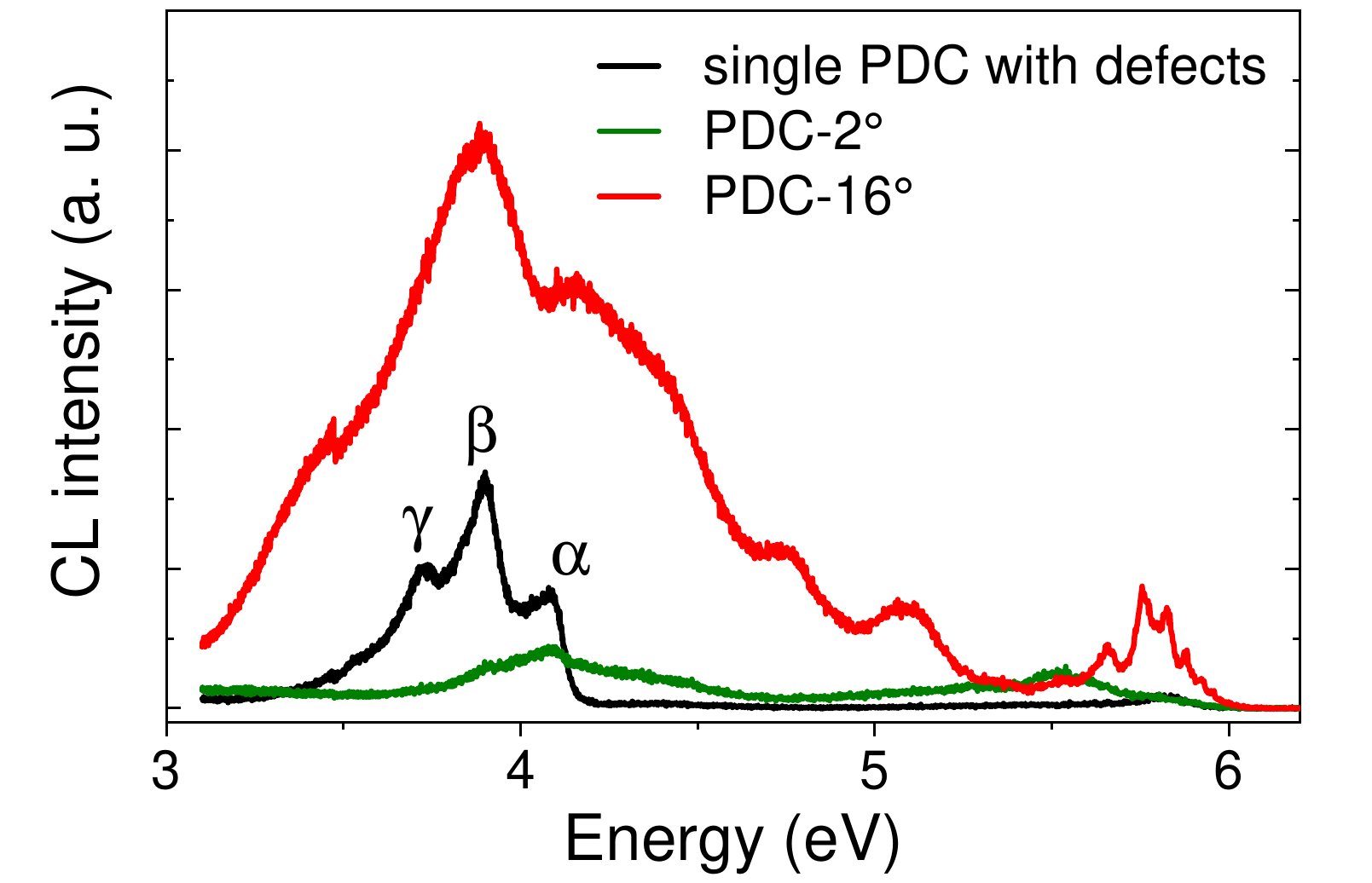}
\caption{CL spectra measured at 300~K on PDC-2°, PDC-16°, and on a single PDC crystal under the same excitation conditions (3~kV, 280~pA).}
\label{F10}
\end{figure}

Figure~\ref{F10} shows spectra measured on a slightly twisted PDC sample in green and a highly twisted PDC sample in red, compared to a spectrum measured on a single PDC crystal in black. The single crystal shows emission from point defects (labeled $\alpha$, $\beta$, $\gamma$) that occur locally in some crystals. As shown in Fig.~\ref{F4}(c) of the main text, the CL intensity of the hBN-hBN interface decreases with decreasing twist angle. For the PDC-2° sample, the intensity of the interface luminescence is lower than the intensity of the native color center emission. To avoid confusion between native defects and interface emission, only highly twisted single crystal structures without significant native defect emission in the 3-6~eV range have been investigated.

\subsection*{Appendix D: Luminescence decay as function of the energy within the broad 4-eVm luminescence.}
Given the linewidth of the emission at 4-eV luminescence, which extends between 3 and 5~eV, the question arises whether it is the result of a population of emitters of different energies or whether the signal is intrinsically broad. To check this, the decay of the luminescence is studied at different energies of the broad emission. 

\begin{figure}[h!]
\includegraphics[scale=0.29]{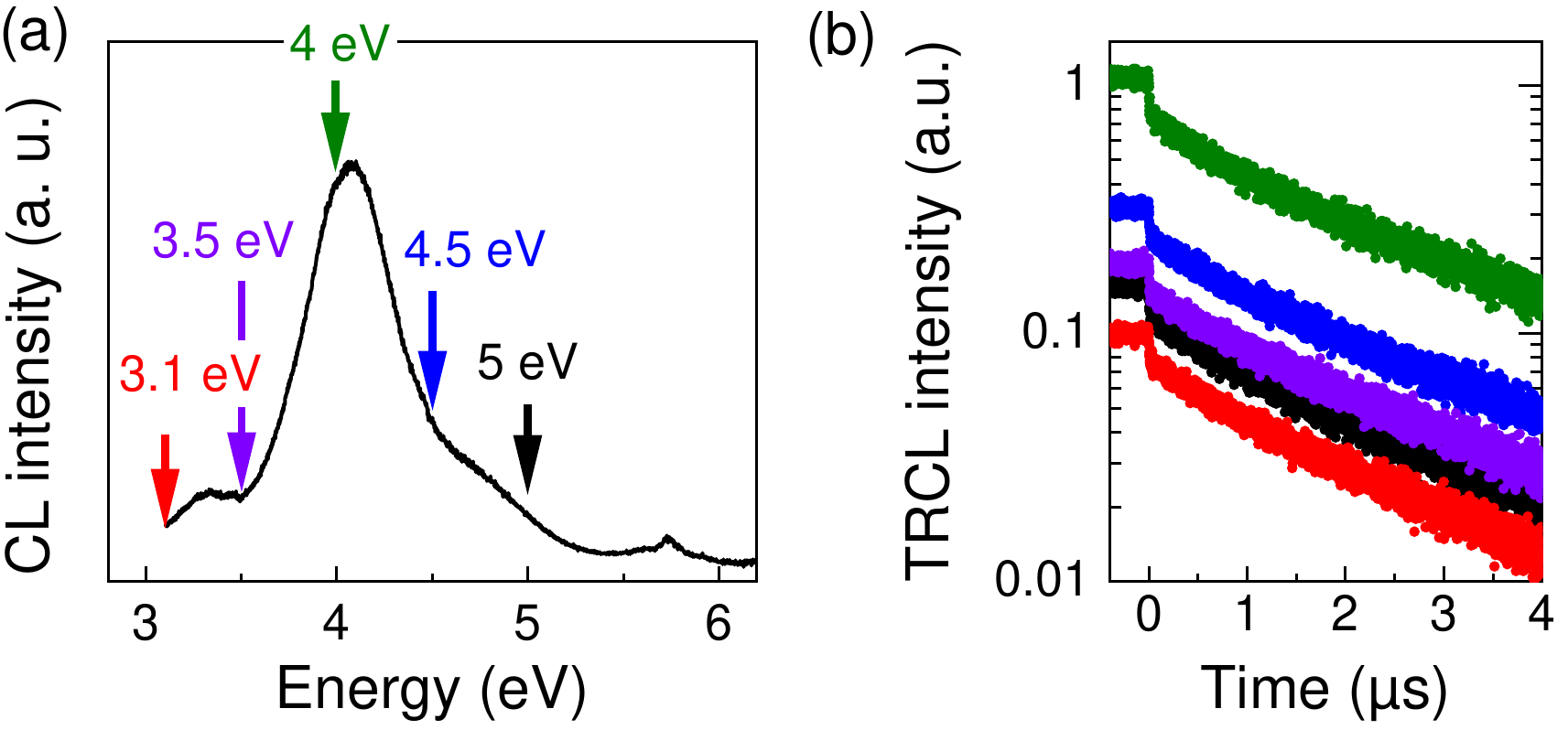}
\caption{CL spectrum (3~kV, 280~pA) measured on the APHT-29° homostructure at 300~K. (b) Decay of the luminescence intensity, filtered at different energies, recorded in TRCL after the excitation was stopped at t = 0. The energies corresponding to each color are indicated by arrows in (a).}
\label{F11}
\end{figure}

Figure~\ref{F11}(a) shows the spectrum of the APHT-29° sample, dominated by the luminescence of interest with little Fabry-Pérot interference contrast. Figure~\ref{F11}(b) shows the TRCL decays of the luminescence filtered at different energies ($\pm$0.15~eV). The decay dynamics are perfectly identical for all investigated energies. This result suggests a unique recombination mechanism for this particularly broad band. The large spectral width of the 4-eV band thus appears to be an intrinsic feature of the luminescence process of twisted hBN-hBN homostructures.

\subsection*{Appendix E: Low temperature spectrum of twisted hBN-hBN homostructure.}
\begin{figure}[h!]
\includegraphics[scale=0.32]{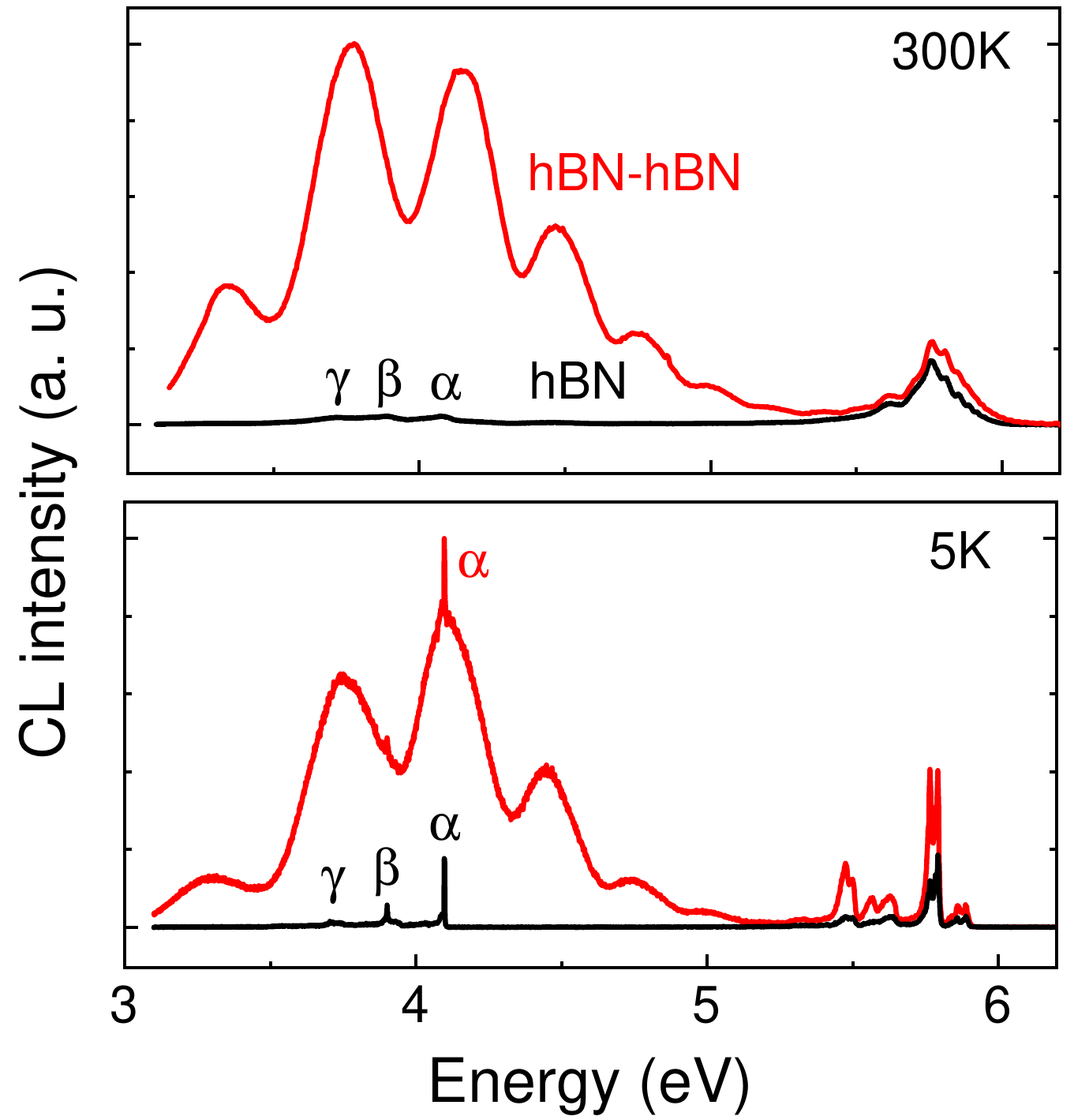}
\caption{CL spectra measured at room and cryogenic temperature on a single hBN flake, in black, and on a hBN-hBN homostructure with a 23° twist, in red, both fabricated from the same HPHT crystal.}
\label{F12}
\end{figure}

Figure~\ref{F12} shows spectra obtained on a 23° twisted hBN-hBN structure and on a hBN single crystal at room temperature and at cryogenic temperature. On the single crystal, the luminescence is dominated by the free exciton luminescence occurring at 5.75~eV, while on the hBN-hBN structure the emission is dominated by the broad luminescence occurring between 3 and 5~eV. The band is modulated by Fabry-Perot interference. We observe that the band remains extremely broad even at cryogenic temperatures. A color center emission also appears on the hBN-hBN structure and on the single hBN crystals, with a zero-phonon band at 4.06~eV ($\alpha$) and its phonon replica ($\beta$ and $\gamma$). This emission shows a large thermal broadening typical of color centers, indicating that its nature is different from that of the hBN-hBN interface emission, whose broadening remains around 2~eV even when the sample holder temperature reaches 5K.

\subsection*{Appendix F: Simulation of TRCL decays with rate equations.}
The TRCL decay of the luminescence intensity is related to the evolution of the exciton populations and thus to their interactions. The scenario of exciton population interactions discussed in the main text is the following: the free exciton population $N_{3D}$, generated in the crystal volume by the primary excitation, are trapped (rate $C$) at the interface. The excitons accumulate at the 2D interface under a dark and long living form ($N_{2D}$ population). They are finally converted into N$_{ST}$ by their self-trapping (rate $ST$) and into N$_{3D}$ by their detrapping (rate $D$). This scenario, illustrated in Fig.~\ref{F7} of the main text, leads to the following differential equations:
\begin{equation}
\left\{
\begin{array}{rcl}
\frac{dN_{3D}}{dt} & = & - \frac{N_{3D}}{\tau_{0}} -C N_{3D}  +D N_{2D} + G(t) \\ \\
\frac{dN_{2D}}{dt} & = & -(D + ST) N_{2D}+C N_{3D} \\ \\
\frac{dN_{ST}}{dt} & = & - \frac{N_{ST}}{\tau_{ST}} +ST N_{2D} \\
\end{array}
\right.
\end{equation}
where $\tau_{0}$ is the lifetime of the free exciton in the volume (without interface) and $\tau_{ST}$ is the lifetime of the X$_{ST}$ excitons. $G$ denotes the generation rate of free excitons, which is constant during continuous excitation and then zero when the excitation stops. A priori, $C$ varies slightly with temperature due to the weak evolution of the exciton diffusion length with temperature \cite{Roux2023}, while the rates of $D$ and $ST$ are thermally activated.

\begin{figure}[h!]
\includegraphics[scale=0.24]{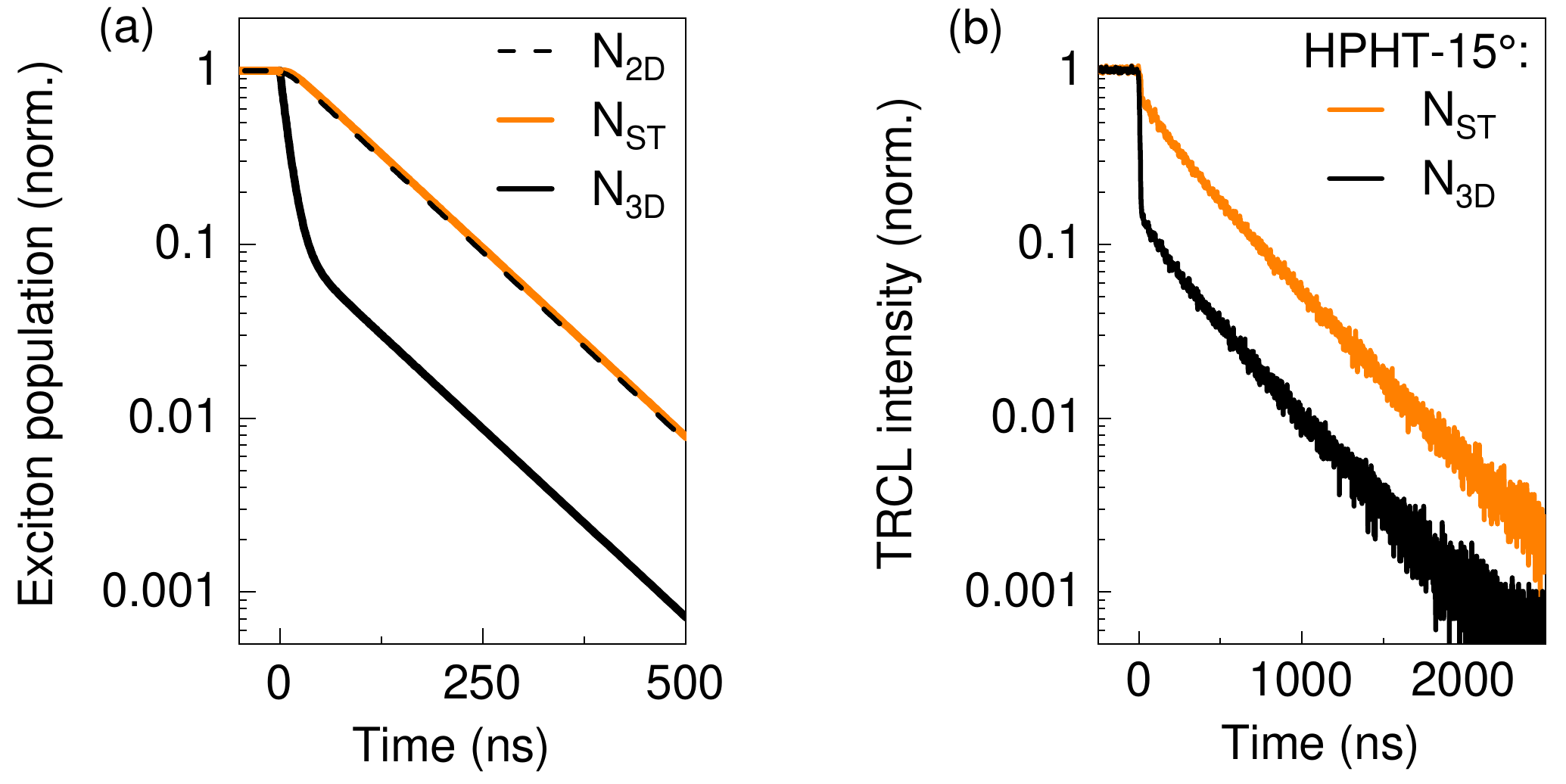}
\caption{(a) Simulated decays of the N$_{2D}$ and N$_{3D}$ exciton populations for $\tau_{s}$=10~ns, $\tau_{l}$=100~ns and A$_{l}$=0.1, and of the N$_{ST}$ self-trapped exciton population for $\tau_{ST}$ = 10~ns. (b) Experimental decays for the N$_{ST}$ and N$_{3D}$ exciton populations on the HPHT-15° structure measured in TRCL at 5~kV, 27~pA, 300~K.}
\label{F13}
\end{figure}

The first two equations of the system are not coupled to N$_{ST}$, they could be considered independently. We find a superposition of two exponential decays for both N$_{3D}$ and N$_{2D}$ exciton populations: one short with a characteristic time $\tau_{s}$ and the other long with a characteristic time $\tau_{l}$ $>$ $\tau_{s}$. These times can be written with A = 1/$\tau_{0}$ + $C$, and B = $D$ + $ST$:
\begin{equation}
\frac{1}{\tau_{s}}=\frac{A+B+ \sqrt{(A-B)^{2}+4CD}}{2}
\end{equation}
\begin{equation}
\frac{1}{\tau_{l}}=\frac{A+B- \sqrt{(A-B)^{2}+4CD}}{2}
\end{equation}
The interruption of the steady excitation at the initial time results in G(t) = 0 for t$>$0. The previously established steady equilibrium imposes the initial conditions:

\begin{equation}
\frac{N_{2D}(0)}{N_{3D}(0)} = \frac{C}{B}
\end{equation}
And the N$_{3D}$ population is written as:
\begin{equation}
\frac{N_{3D}(t)}{N_{3D}(0)} = A_{l}e^{-\frac{t}{\tau_{l}}} + (1-A_{l}) e^{-\frac{t}{\tau_{s}}} 
\end{equation}
With the weight of the long exponential, A$_{l}$, given by:
\begin{equation}
A_{l} = \frac{\tau_{l}}{\tau_{l}-\tau_{s}} \left( 1 - \frac{1}{B\tau_{l}} \right) \approx 1 - \frac{1}{B\tau_{l}}
\end{equation}
This gives:
\begin{equation}
B = \frac{1}{\tau_{l}} \frac{1}{1-A_{l}}
\label{eq:5.8}
\end{equation}
The parameter B=$D$+$ST$ can therefore be extracted from the TRCL decays of the free exciton luminescence from the previous equation, or from the simplified form when A$_{l}$ $\ll$ 1:
\begin{equation}
\frac{1}{\tau_{l} \simeq D+ST  }
\end{equation}
Calculations show that all three populations follow the same decay time, as can be seen in Fig.~\ref{F13}(a), which is consistent with the experimentally observed decays of the N$_{ST}$ and N$_{3D}$ populations in Fig.~\ref{F13}(a) (see also Fig.~\ref{F3}(b) in the main text). The long component of the TRCL decay being governed by the detrapping and self-trapping rates of the N$_{2D}$ population that occurs in the long term.

%

\end{document}